\documentclass{birkjour}

\catcode`\@=11
\def\newsymbol#1#2#3#4#5{\let\next@\relax%
 \ifnum#2=\@ne\else%
 \ifnum#2=\tw@\let\next@\msyfam@\fi\fi%
 \mathchardef#1="#3\next@#4#5}
\def\mathhexbox@#1#2#3{\relax%
 \ifmmode\mathpalette{} {\m@th\mnnathchar"#1#2#3}
 \else\leavevmode\hbox{$\m@th\mathchar"#1#2#3$}\fi}
\font\tenmsy=msbm10
\font\sevenmsy=msbm7
\font\fivemsy=msbm5
\newfam\msyfam
\textfont\msyfam=\tenmsy
\scriptfont\msyfam=\sevenmsy
\scriptscriptfont\msyfam=\fivemsy
\edef\msyfam@{\hexnumber@\msyfam}

\catcode`\@=\active
\newtheorem{theorem}{Theorem}[section]
\newtheorem{proposition}[theorem]{Proposition}
\newtheorem{lemma}[theorem]{Lemma}
\newtheorem{corollary}[theorem]{Corollary}
\newtheorem{definition}[theorem]{Definition}
\newtheorem{example}[theorem]{Example}
\newtheorem{remark}[theorem]{Remark}

\newtheorem{assumption}[theorem]{Assumption}

\newcommand{\bm}[1]{\mathcal #1}

\newcommand{\eq}[1]{\begin{equation}\label{#1}}
\newcommand{\en}{\end{equation}}
\newcommand{\eqn}{\begin{eqnarray*}}
\newcommand{\enn}{\end{eqnarray*}}
\newcommand{\eqnn}{\begin{eqnarray}}
\newcommand{\ennn}{\end{eqnarray}}
\renewcommand{\proof}{{\noindent \it Proof:\ }}
\renewcommand{\qed}{\hfill $\Box$ \par\medskip}
\newcommand{\BR}{{\RR^d}}

\newcommand{\TTT}[1]{{\bf (#1)}}
\newcommand{\HS}{{H_V}}
\newcommand{\td}{T_{\rm d}}%\rm d}}
\newcommand{\D}{{\rm D}}

\newcommand{\eps}{\varepsilon}
\newcommand{\bi}{\begin{itemize}}
\newcommand{\ei}{\end{itemize}}
\newcommand{\bd}[1]{\begin{definition}\label{#1}}
\newcommand{\ed}{\end{definition}}

\newcommand{\CC}{{\mathbb{C}}}
\newcommand{\RR}{{\mathbb{R}}}
\newcommand{\bl}[1]{\begin{lemma}\label{#1}}
\newcommand{\el}{\end{lemma}}
\newcommand{\bc}[1]{\begin{corollary}\label{#1}}
\newcommand{\ec}{\end{corollary}}
\newcommand{\bt}[1]{\begin{theorem}\label{#1}}
\newcommand{\et}{\end{theorem}}
\newcommand{\bp}[1]{\begin{proposition}\label{#1}}
\newcommand{\ep}{\end{proposition}}

\newcommand{\kak}[1]{(\ref{#1})}
\newcommand{\LR}{{L^2(\BR)}}

\newcommand{\ms}[1]{\mathcal{#1}}

\newcommand{\f}{^{-1}}
\newcommand{\lk}{\left(}
\newcommand{\rk}{\right)}

\newcommand{\MSD}{\mathcal D}
\newcommand{\MSK}{\mathcal K}

\newcommand{\mft}{{\mathfrak{t}}}

\newcommand{\add}{a^{\ast}}

\newcommand{\ov}[1]{\overline{#1}}

\newcommand{\half}{\frac{1}{2}}
\newcommand{\han}{{\half}}

\def\bbbone{{\mathchoice {\rm 1\mskip-4mu l} {\rm 1\mskip-4mu l}
{\rm 1\mskip-4.5mu l} {\rm 1\mskip-5mu l}}}
\def\one{\bbbone}

\newcommand{\s}{\sigma}%{\rm Sp}}
\newcommand{\hhh}{{\mathcal H}}

\renewcommand{\d}{\displaystyle}

\makeatletter
\@addtoreset{equation}{section}
\makeatother
\def\theequation{\arabic{section}.\arabic{equation}}

\begin{document}

\title{Ultra-Weak Time Operators of   Schr\"odinger Operators}

%----------Author 1
\author{Asao Arai}
\address{%
Department  of Mathematics,\\ Hokkaido University, \\ Sapporo 060-0810,\\ Japan. }
\email{arai@ math.sci.hokudai.ac.jp}

%----------Author 2 

\author{Fumio Hiroshima} %
\address{Faculty of Mathematics,\br Kyushu University,\br Fukuoka 819-0395,\br Japan}
\email{hiroshima@math.kyuhsu-u.ac.jp}

\thanks{This research was supported by KAKENHI 15K04888 from JSPS (A.A.),  partially supported by CREST, JST, 
and Challenging Exploratory Research 15K13445 from JSPS (F.H)}

\subjclass{}%Primary 99Z99; Secondary 00A00}

\keywords{Time operators, Weyl relations,  CCR,  Schr\"odinger operators}

\date{\today}

\begin{abstract}
In an abstract framework, a new concept on time operator, {\it ultra-weak time operator}, is introduced,
which is a concept weaker than that of weak time operator.
Theorems on the existence of an ultra-weak time operator
are established.
As an application of the theorems,  it is shown that  Schr\"odinger
operators $\HS$ with potentials $V$ obeying suitable conditions, including the Hamiltonian of the hydrogen atom, 
have  ultra-weak time operators. Moreover,  a class of Borel measurable 
functions $f:\RR\to\RR$ such that $f(\HS)$ has an ultra-weak time operator is found. 
\end{abstract}

%%% ----------------------------------------------------------------------
\maketitle
%%% ----------------------------------------------------------------------
\tableofcontents

\section{Introduction}

The present  paper  concerns  a time operator in quantum theory which is
defined, in a first stage of cognition,  as a symmetric operator canonically conjugate  to a Hamiltonian if it exists.
The uncertainty relation which is derived from the canonical commutation relation of a time operator and a Hamiltonian may be
interpreted as a mathematically rigorous form of {\em time-energy uncertainty relation}. 
Moreover time operators may play important  roles in quantum phenomena \cite{miy01,ara05,ara08b,am08b,mme08,mme09}. 
To explain motivations for studying time operators,  we begin with a brief historical review on  time and time operator
in quantum theory (cf. also \cite[Chapter 1]{mme08}).

\subsection{Historical backgrounds}

In the old quantum theory, N. Bohr assumed  that the interaction of the electrons 
in an atom with   an electromagnetic field  causes transitions among the allowed electron  orbits in such a way that the transitions are accompanied by the absorption or the emission of electromagnetic radiations by the atom. 
In this hypothetical theory, however,
no principle for the timing  of occurrence of these transitions  was  shown.   
The new quantum theory presented
by Heisenberg (1925), Born-Heisenberg-Jordan (1926) and Schr\"odinger (1926)
provides a method of calculating the transition probabilities, but the question of the timing at which
the events occur was not addressed explicitly.

Heisenberg introduced two kinds of uncertainty relations, i.e.,  the uncertainty relation for position and momentum, 
and that  for time and energy.  
He argued  (\cite[p.179, equation (2)]{hei27}) that the imprecision within which the instant of transition is specifiable is given by $\Delta t$  obeying
the uncertainty relation
\eq{uto}
(\Delta t) (\Delta E)\sim \hbar=\frac {h}{2\pi}
\en
with the change $\Delta E$ of energy in the quantum jump, where $h$ is the  Planck constant. 
Although many of the issues involved in the uncertainty  principle for position and momentum have been clarified so far, 
similar clarity has not yet been achieved on the uncertainty principle for  time and energy.
For example, in \cite{aa90,mt45},  uncertainty relation \kak{uto}
is derived, but $\Delta t$ is not considered an imprecision of measurement on time; 
 interpretations such as \lq\lq{a measurement act of the time gives an unexpected change to an energy level}" or \lq\lq{it dishevels a clock to have been going to measure energy exactly}" may be invalid unless any restrictions are imposed
depending on  measurement setups.
In addition,  the definition of $\Delta t$ seems to  vary  from case to case. 

\subsection{Time in quantum theory and time operator}\label{subsec-time}

It is said that there exists a three-fold role of 
time in quantum theory ( see, e.g., \cite{bus01} and \cite[Chapter 3]{mme08}).  
Firstly time is identified as the parameter entering the Schr\"odinger equation, 
which is a differential equation  describing 
 the causal continuous 
change of  states of a quantum system,  
and measured by a laboratory clock. Time in this sense is called the  external time. 
The external time measurement is  carried out with clocks that are not dynamically connected with objects investigated in  experiments. 

By contrast, time as a dynamical one can be  defined by 
the dynamical behavior of quantum objects. A dynamical time is defined and measured in terms of a physical system undergoing changes. Examples include
the  linear uniform  motion of a free particle and the oscillation of the atoms   in an atomic clock. 

Finally,  time can be considered as a quantum object  which  forms a  canonical pair with a Hamiltonian in a suitable sense.
As already mentioned, time in this sense is called a time operator in its
simplest form. There is in fact  a hierarchy of time operators as is shown below.
The main purpose of the present paper is to analyze this hierarchy mathematically and 
to establish  abstract existence theorems on time operators
in relation to  the hierarchy with applications to Schr\"odinger Hamiltonians. 

A simple example of  time operator is given as follows.
A non-relativistic quantum particle with mass $m>0$ under the action of a constant force $F\in \RR\setminus\{0\}$
in the one-dimensional space $\RR$ 
is governed by the Hamiltonian 
$$
H_F=\frac{1}{2m} P^2-FQ
$$ 
acting in 
$L^2(\RR)$, the Hilbert space of square integrable Borel measurable  functions on $\RR$,  with the momentum operator $P=-iD_x$ ($D_x$ is the generalized differential operator in the variable $x\in\RR$)\footnote{We use the physical unit system where $\hbar=1$.} and
$Q$  being the multiplication operator by $x$.
It is shown that
$H_F$ is essentially self-adjoint on $C_0^{\infty}(\RR)$, the space of
infinitely differentiable functions on $\RR$ with compact support,  and hence its closure $\overline{H}_F$ is self-adjoint
(but, note that $\overline{H}_F$ is neither bounded from below nor from  above).
The self-adjoint operator 
$T_F=P/F $ 
satisfies
the canonical commutation relation (CCR) 
$$
[H_F,T_F]=-i\one
$$
on a dense domain $\MSD$ (e.g., $\MSD=C_0^{\infty}(\RR)$),  
where $\one$ denotes identity and $[A,B]:=AB-BA$.
This shows that $T_F$ is a canonical conjugate operator to the Hamiltonian $H_F$ and hence a time operator of $H_F$. 
From the CCR, one can derive 
the uncertainty relation
of Heisenberg type 
$$
(\Delta H_F)_{\psi}(\Delta T_F)_{\psi}\geq \half
$$
for all unit vectors $\psi\in \MSD$, where $(\Delta A)_{\psi}$
denotes the uncertainty  of $A$ with respect to the state vector $\psi$
(see (\ref{Delta-A}) for its definition). This inequality may be interpreted as a form of time-energy uncertainty relation
in the present model. 

As for time  operator, however, there is a long history of confusion and controversy\footnote{A germ of the notion of time
operator is found already in Heisenberg's paper \cite[pp.177--179]{hei27} in 1927.}.  
The origin of this may come from the statement of  Pauli made in 1933 (\cite[p.63, footnote 2]{pau33})  that 
the introduction of a time observable $T$ satisfying the CCR
\eq{to}
[H, T]=-i\one
\en  
with a self-adjoint operator $H$ having a discrete eigenvalue
is basically forbidden. Although there are no explicit arguments  for this statement in the cited literature
(only reference to  Dirac's textbook),  a formal (false in fact) argument leading to the statement may be as follows:
let $\phi$ be an eigenvector of $H$ with a discrete eigenvalue $E$:$H\phi=E\phi$.
Then, using (\ref{to}) formally, one obtains 
$He^{i\eps T}\phi=(E+\eps)e^{i\eps T}\phi\cdots (*)$ for all $\eps\in\RR$.
Hence $e^{i\eps T}\phi$ is an eigenvector of $H$ with eigenvalue $E+\eps$.
Since $\eps\in\RR$ is arbitrary, it follows that each point in $\RR$ is an eigenvalue of $H$.
But this  obviously contradicts the discreteness of  eigenvalues of $H$. 
It should be  noted, however, 
that this  argument is very  formal, in particular, no attention  was paid to the domain of the operators involved. E.g.,
if $\phi$ is not in the domain of $T^n$ for some $n\in \mathbb{N}$, then 
the expansion $e^{i\eps T}\phi=\sum_{n=0}^{\infty}(i\eps)^nT^n\phi/n!$  is meaningless; even in the case
where $\phi$ is in the domain of $T^n$ for all $n\in\mathbb{N}$,
$\sum_{n=0}^{\infty}(i\eps)^nT^n\phi/n!$ is not necessarily  convergent ; moreover, 
$e^{i\eps T}\phi$ is not necessarily in the domain of $H$ and, if $e^{i\eps T}\phi$ is not in the domain of $H$, then
 $(*)$ is meaningless.
 
It is well known \cite[p.2]{pu67} that at least one of $T$ and $H$ satisfying the CCR (\ref{to}) on a dense domain is an unbounded operator and, for unbounded operators,  their domain must be carefully considered.
As a matter of fact,  the above  argument is incorrect and so is the Pauli's statement too. 
Indeed, one can construct a self-adjoint operator $H$ which is bounded from below with  purely discrete spectrum and a self-adjoint operator $T$ such that \kak{to} holds
on a dense domain. This was pointed out in \cite[p.4]{bus01} and mathematically rigorous constructions
of such time operators $T$  have been done in \cite{gal02,am08b}.

The history of studies on time operators as well as on representations of CCR
suggests that there may be a hierarchy of time operators and this indeed is the case  as is shown below in the present paper. 
It is important to distinguish each class from the others in the  hierarchy.
In our words,  the time observable $T$ such that the above formal argument
may take a rigorous form  is  an {\em ultra-strong} time operator (see Remark \ref{rem-u-strong}  below), since the {\em operator equality} $e^{-i\eps T}He^{i\eps T}=H+\eps,\,\eps\in\RR\cdots (\dagger)$
is tacitly assumed in the above argument in fact, which, however, 
is {\em not equivalent} to \kak{to} in the mathematically rigorous sense \cite{fug67},
and, if $H$ is self-adjoint, then $(\dagger)$ is equivalent to the Weyl relation $e^{i\eps T}e^{itH}=e^{-it\eps}e^{itH}e^{i\eps T},\,t,\eps\in\RR\cdots (\dagger\dagger)$, meaning that $T$ is an ultra-strong  time operator of $H$  (but, if 
$H$ is not essentially self-adjoint, then $(\dagger)$ does not imply $(\dagger\dagger)$ with $H$ replaced by the closure
$\overline{H}$ of $H$, because, in this case, \lq\lq{\,$e^{it\overline{H}}$\,}" is meaningless as a unitary operator).
If a self-adjoint operator $H$ has an ultra-strong time operator $T$,  then
$T$ is a strong time operator of $H$ (see (\ref{hierarchy}) below) and hence 
$H$ is  absolutely continuous (see Proposition \ref{time} below) so that  $H$ has no eigenvalues. Therefore,
in this case, the above argument becomes meaningless.
Moreover, if $H$ is semi-bounded, then no strong time operator $T$ of $H$ is essentially self-adjoint 
(\cite{miy01}, \cite[Theorem 2.8]{ara05}) and
hence \lq\lq{\,$e^{i\eps \overline{T}}$\,}" makes no sense as a unitary operator. In this sense too, 
the above argument is meaningless. 

It has been absurd that studies  on time observables have been ruled out for so many years due to the 
Pauli's statement without any questions. If one could have carefully examined the Pauli's statement
with mathematically rigorous thinking, then one could have found incorrectness of it.

\subsection{Rough description of main results}

As already mentioned, a time operator $T$ of a self-adjoint operator $H$ is 
defined  to be a symmetric operator  satisfying CCR~\kak{to} on a suitable dense domain (we shall 
give a more detailed description of time operators in Section \ref{sec2}). 
Another  approach to consider time operators as observables is an application of   
positive operator valued measures (POVM) \cite[Chapter 10]{mme08}. In this paper,  however, 
we take an operator-theoretical approach to classify time operators 
 and to construct a time operator 
for a given self-adjoint operator without invoking POVM. 
Consequently we are led to extend  the conventional notion of time operator.
Indeed, commutation relation \kak{to} can be weakened in at least two manners
and we find  a time  operator $T$  for  each weakened form.  
As we have learned  from the formal argument on the Pauli's statement,  
taking care of domains of $T$ and $H$ is crucial not to be led to incorrect
conclusions. Thus 
the domain of time operators  is one key ingredient to study them. 

We now outline main results obtained in  the present paper in (1)--(5) below (rigorous statements of assumptions and results will be given from Section \ref{sec2}).
Let $H$ be a self-adjoint operator acting in a complex Hilbert space $\hhh$.  

\medskip

\noindent
{\bf (1)  Ultra-weak time operators and   a hierarchy of time operators}.
It has so far been known that there are at least three classes of time operators \cite{ara08b,ara08c}, i.e.,
 time operators as canonical conjugates of a Hamiltonian in the conventional  sense, which may be  called 
{\em ordinary time operators} to distinguish
them from other classes of time operators, {\em strong time operators} and {\em weak time operators}. 
In the present paper, in addition to these classes of time operators,  
we introduce   a new concept on time operator, which we call {\it ultra-weak time operator},   
and study it. An ultra-weak time operator, however, is  not an operator in general, but defined to be a
sesquilinear form $\mft:\MSD_1\times \MSD_2\to\CC$ with non-zero subspaces   $\MSD_1$ and $\MSD_2$ of $\hhh$ such that 
$$
\mft[ H\phi,\psi] -\mft[H\psi,\phi]^*=-i(\phi,\psi),\quad \psi,\phi\in\ms E,
$$
where, for $z\in\CC$, $z^*$ denotes the complex conjugate of $z$,  $(\,\, ,\,\,)$ is the inner product of $\hhh$ (linear in the second variable) and
$\ms E$ is a non-zero subspace of $\hhh$ (for the rigorous definition of $\mft$, see Definition \ref{def-uwt}). 
The class of ultra-weak time operators may be compared to
the  space of distributions in the context of theory of functions (as there exists a distribution which is not a function,
there may  exist an ultra-weak time operator which is not an operator).

For convenience, we also introduce the concept of ultra-strong time operator which has been already mentioned above.
These five classes of time operators form  a hierarchy in the following sense:
\begin{align}
\{\mbox{ultra-strong t.o.}\}\!\subset\! \{\mbox{strong t.o.}\}
\!&\subset\! \{\mbox{t.o.}\}\!\nonumber\\
&\subset\! \{\mbox{weak t.o.}\}\!\subset\!
\{\mbox{ultra-weak t.o.}\}, \label{hierarchy}
\end{align}
where t.o. is abbreviation of \lq\lq{time operators}".
See Section \ref{sec2} below for more details.
Generally speaking, it is expected that each class in the hierarchy of time operators  
has proper roles in connection with quantum phenomena. 
In this paper, we particularly concentrate our attention on  
strong time operators, time operators and ultra-weak time operators.
As a possible physical aspect of ultra-weak time operators, a weak form of  uncertainty relation
is given (see Proposition~\ref{U1}).

\medskip
 
\noindent
{\bf (2) Existence of strong time operators in an abstract framework}. 
A strong time operator $T$ of a self-adjoint operator $H$ is defined through 
the  weak Weyl relation (see Definitions~\ref{wwr} and \ref{1100}). 
It is known that \kak{to} is satisfied on a dense domain and the spectrum 
$\s(H)$ of $H$ must be purely absolutely continuous. Hence, if
$H$ has an eigenvalue, no strong time operator of $H$ exists.  
Then a natural question is to ask the existence of a strong time operator for an absolutely continuous self-adjoint operator. 
We introduce a class $S_0(\hhh)$ of self-adjoint operators on $\hhh$ in Definition \ref{S1} and prove the following theorem (Theorem \ref{thm-ac-st}):
\begin{theorem}\label{M0} Assume that $\hhh$ is separable and that $H\in S_0(\hhh)$. 
Then $H$ has  a strong time operator.
\end{theorem}

It may be interesting  to  consider extensions of this theorem to a more general class of absolutely continuous
self-adjoint operators. But this will be done elsewhere.
In this paper, we
next proceed to construction of  a time operator for a self-adjoint  operator 
which has point spectra (eigenvalues).

\medskip

\noindent
{\bf (3) Existence of time operators of  a self-adjoint operator with point spectra}.
As for general existence of time operators of a self-adjoint operator $H$ with
point spectra, only  limited classes of $H$
have been found \cite{gal02,am08a,am08b,ara09}.
In this paper we extend these results (Theorem \ref{thm-d-top}): 
 
\begin{theorem}\label{M1}
Let $\s (H)=\{E_n\}_{n=1}^\infty$,  
$E_1<E_2<\cdots $ and $\lim_{n\to\infty}E_n=\infty$.
Then there exists a  time operator $T$ of  $H$.
\et

In \cite{gal02,am08a,am08b}, time operators of $H$ having purely discrete spectrum are constructed, but 
the growth condition $\sum_{n=1}^\infty{1}/{E_n^2}<\infty$  for $\{E_n\}_n$ is imposed. This condition seems to be artificial.  An important point 
in Theorem \ref{M1} is that  this condition is not required.  
We show in Subsection \ref{subsec-III} that
the noncommutative harmonic oscillator Hamiltonian \cite{iw07} and the Rabi Hamiltonian 
\cite{rab36,rab37,bra11,mps14} are included  in this class  of Hamiltonians as concrete examples.

\begin{remark}{\rm After submitting the first version of the present paper, we have learned
that Teranishi \cite{te16} has  proved a theorem essentially same as Theorem \ref{M1} by
a method different from ours.}
\end{remark}

\medskip

\noindent
{\bf (4) Ultra-weak time operators}.
We  also establish a theorem on the existence of ultra-weak time operators 
for a general class of self-adjoint operators with infinitely many  discrete eigenvalues but  the accumulation point is not $\infty$ (Theorem \ref{thm-uwt}). 
\begin{theorem}
Suppose  that 
 $\s (H)\setminus\{0\}=\{E_j\}_{j=1}^\infty$,  
$E_1<E_2<\cdots <0$, $\lim_{j\to\infty}E_j=0$,  and $0$ is not an eigenvalue of $H$.
Then there exists an ultra-weak  time operator $\mft$ of  
$H$.
\end{theorem}

It will be seen in Subsection \ref{4.1} 
that $\mft[\phi, \psi]=(\phi,A\psi)$  formally with some operator $A$. 
The crucial point is that $A$ is of the form 
$$A=-\half(T_{-1}H^{-2}+H^{-2}T_{-1}),$$
where $T_{-1}$ denotes a time operator of $H^{-1}$.  
It is difficult to show, however, that $\D(A)\not=\{0\}$ and $\D(HA)\cap \D(AH)\not=\{0\}$. 
This is the reason why the introduction of an ultra-weak time operator $\mft$ 
as a sesquilinear form is needed  and may be even  natural.

\medskip

\noindent
{\bf (5) Ultra-weak time operators for Schr\"odinger operators}.
Finally, by applying the results described  in (1)--(4) above, we  construct  an ultra-weak time operator for 
a class of Schr\"odinger operators, including the Hamiltonian of the hydrogen atom.
It is shown in Theorem \ref{main2} that, for
a class of potentials $V:\RR^d\to\RR$,
the $d$-dimensional Schr\"odinger operator
$$
\HS:=-\frac 1{2m}\Delta+V
$$
acting in $L^2(\RR^d)$ has an ultra-weak time operator, 
where $\Delta$ is the $d$-dimensional generalized Laplacian.
Below are  some examples of $\HS$ having an ultra-weak time operator (see Subsection \ref{ex}
for more details).

(i) Let $U\in L^\infty (\RR^3)$ and  
$$V (x):=\frac{U (x)}{ (1+|x|^2)^{\han +\eps}}.$$ 
Suppose that 
$U$  is negative,  continuous,  spherically symmetric and 
satisfies that 
$U (x)=-1/|x|^\alpha$ for $|x|>R$  
with $0<\alpha<1$ and $R>0$. 
For each $\alpha$,  we 
can choose $\eps>0$ such that 
$2\eps+\alpha<1$. 
Then
$H_V$ has an ultra-weak time operator.
See Example \ref{47}. 

(ii) Let  
$$H_{\rm hyd}:=-\frac 1{2m}\Delta-\frac{\gamma}{|x|}$$ be 
the $3$-dimensional hydrogen Schr\"odinger operator with a constant $\gamma> 0$. Then 
$H_{\rm hyd}$
has an ultra-weak time operator. See Example \ref{ex-hyd}. 

(iii)
Suppose that $\HS$ has an ultra-weak time operator. 
Then, under some conditions, we can 
show that the following operators  $f(\HS)$ also have an ultra-weak time operator (see Theorem \ref{thm-uwt-HS}):
\bi
\item[(a)]
$f (\HS)=e^{-\beta \HS}$ for $\beta\in\RR\setminus\{0\}$;
\item[(b)]
$f (\HS)=\sum_{j=0}^N a_j H_V^j$ ($a_j\in\RR, N\in{\mathbb N}$);
\item[(c)]
$f (\HS)=\sin (2\pi \beta \HS)$ for $\beta\in\RR\setminus\{k/2E_j|k\in{\mathbb Z}, j\in{\mathbb N}\}$, where $\{E_j\}_{j\in{\mathbb N}}$ denotes the discrete spectrum of $\HS$.
\ei
See Examples \ref{exp}, \ref{poly} and \ref{411}. 

In the next section we give definitions of terminology used in this paper and remarks 
from mathematical point of view.

\section{Mathematical Backgrounds of Time Operators}\label{sec2}

\subsection{A review on mathematical analysis on time operators}

Mathematical analysis on time operators has been  developed
in the  papers \cite{miy01,gal02,gcb04,ara05, ara07,ara08a,ara08b,ara08c,am08a,am08b,ara09,hkm09}. 
Let $A$ and $B$ be linear operators on  a complex  Hilbert space $\hhh$,  satisfying 
the canonical commutation relation
\begin{equation}
[A, B]=-i\one
\label{CCR}
\end{equation}
on a non-zero subspace $\MSD\subset \D(AB) \cap \D(BA)$, where, for a linear operator $L$ on $\hhh$, $\D(L)$ denotes the domain
of $L$.  We call $\MSD$ a {\it CCR-domain} for the pair $(A,B)$.
It is well known \cite[p.2]{pu67} that, if $\MSD$ is  dense in $\hhh$, 
then (\ref{CCR})
 implies that $\hhh$ has to be  infinite dimensional and at least one of  $A$  and  $B$ is  unbounded. We call this property the {\it unbounded property} of CCR. 
It is easy to see that, if $\MSD$ is an invariant subspace of $A$ and $B$, then
$\MSD$ has to be infinite dimensional and hence at least one of $A$ and $B$ as linear operators on $\overline{\MSD}$
(the closure of $\MSD$) with domain $\MSD$ is unbounded.
From representation theoretic point of view,
 $(\hhh, \MSD, \{A,B\})$ is called a {\it representation of the CCR with one degree of freedom} (usually $\MSD$ is assumed to be
a dense invariant subspace of $A$ and $B$, but, here, we do not require this property).

W denote by $(f,g)_{\hhh}$ ($f,g\in \hhh$)  and $\|\cdot \|_{\hhh}$ 
the scalar (inner)  product of  $\hhh$,  linear in $g$ and antilinear in $f$, and the norm of  $\hhh$ respectively. 
But we sometimes omit the subscript \lq\lq{$\hhh$}" in  $(f,g)_{\hhh}$ and  $\|\cdot\|_{\hhh}$ if there is no
danger of confusions. 

The CCR (\ref{CCR}) implies an physically important inequality:  
if $A$ and $B$ in (\ref{CCR}) are symmetric operators on $\hhh$, then (\ref{CCR}) yields the {\it uncertainty relation}
of Heisenberg type \cite[Chapter III, \S 4]{vn32}: 
\begin{equation} 
(\Delta A)_{\psi}(\Delta B)_{\psi}\geq \half\label{UR} 
\end{equation}  
for all $\psi\in \MSD$ with $\|\psi\|=1$, where 
\begin{equation}
(\Delta A)_{\psi}:=\|(A-(\psi, A\psi)_\hhh)\psi\|_\hhh,\,\psi\in\D(A), \,\, \|\psi\|=1, \label{Delta-A}
\end{equation}
the {\it uncertainty} of $A$ with respect to 
$\psi$.\footnote{Inequality (\ref{UR}) can be derived also from a weak version of (\ref{CCR}):
$(A\psi,B\phi)-(B\psi,A\phi)=-i(\psi,\phi),\,\psi,\phi \in \MSD_{\rm w}$, where $\MSD_{\rm w}$ is a non-zero
subspace of $\D(A)\cap\D(B)$.} 

The concept of representation of the CCR with one degree of freedom 
can be extended to the case of finite degrees of freedom.
Let $A_j$ and $B_j$ ($j, k=1,\ldots, d, d\in \mathbb{N}$) be symmetric operators on $\hhh$ and $\MSD$ be
a non-zero subspace of $\hhh$ such that $\MSD \subset \cap_{j,k=1}^d[\D(A_jB_k)\cap \D(B_kA_j)\cap \D(A_jA_k)\cap \D(B_jB_k)]$.
Then the triple $(\hhh,\MSD, \{A_j,B_j|j=1,\ldots,d\})$
 is called a {\it representation of the CCR's with $d$ degrees of freedom}
 if the CCR's with $d$ degrees of freedom 
\begin{equation}
[A_j,B_k]=-i\delta_{jk}\one,\quad [A_j,A_k]=0,\quad [B_j,B_k]=0,\quad j,k=1,\ldots,d\label{CCR-d}
\end{equation}
hold on $\MSD$, where $\delta_{jk}$ is the Kronecker delta. The subspace $\MSD$ is called a {\it CCR-domain}
for $\{A_j,B_j|j=1,\ldots,d\}$.

There is a stronger version of representation of the CCR's with $d$ degrees of freedom.
A set $\{A_j,B_j|j=1,\ldots,d\}$ of self-adjoint operators on $\hhh$ is
called a {\it Weyl representation of the CCR's with $d$ degrees of freedom}  
 if the {\it Weyl relations}
\begin{equation}
e^{-is A_j} e^{-itB_{k}}=e^{ist\delta_{jk}}e^{-itB_{k}}e^{-isA_j},\quad j,k=1,\ldots,d,\, s,t\in\RR\label{Weyl}
\end{equation}
hold. 

The  Weyl relations (\ref{Weyl}) imply that there exists a dense invariant domain $\MSD$ of $A_j$ and $B_j$
 ($j=1,\ldots,d$)  such that  (\ref{CCR-d}) holds on  $\MSD$ \cite[Theorem 4.9.1]{pu67}.
Hence the Weyl representation $\{A_j,B_j|j=1,\ldots,d\}$
is a representation of the CCR's with $d$ degrees of freedom.
But {\it the converse is  not true} (e.g., \cite{fug67,sch83b,ara98}).

 A Weyl representation    $ \{A_j,  B_{k}| j, k=1,\ldots, d\}$ of the CCR's with $d$ degrees of freedom  
is said to be {\it irreducible} if any subspace $\MSD$ of $\hhh$ left invariant
by $e^{-itA_j}$ and $e^{-itB_j}$ for all $t\in \RR$ and $j=1,\ldots,d$ is $\{0\}$ or $\hhh$.

In quantum mechanics on the $d$-dimensional space 
$$\BR=\{x=(x_1,\ldots,x_d)|x_j\in \RR\},$$
the momentum operator $P:=(P_1,\ldots,P_d)$ and the position operator $Q:=(Q_1,\ldots,Q_d)$
are defined by $P_j:=-iD_j$ ($D_j$ is the generalized partial differential operator 
in $x_j$)  and 
$Q_j:=M_{x_j} (\mbox{the multiplication operator by }x_j)$, \, $j=1,\ldots, d$.
For all $j=1,\ldots,d$, $P_j$ and $Q_j$ are 
self-adjoint operators on the Hilbert space $\LR$,  satisfying  the CCR's with $d$ degrees of freedom: 
\eq{ccr2}
[P_j,  Q_{k}]=-i\delta_{jk}\one, \quad 
[P_j,  P_{k}]=0, \quad 
[Q_j,  Q_{k}]=0
\en
on the domain $\cap_{j,k=1}^d[\D (P_jQ_{k})\cap \D (Q_{k}P_j) \cap \D(Q_jQ_{k})\cap \D(Q_{k}Q_j) ]$.
Hence $(\LR$, $C_0^{\infty}(\BR)$, $\{P_j,Q_j|j=1,\ldots,d\})$
is a representation of the CCR's with $d$ degrees of freedom, where $C_0^{\infty}(\BR)$ is
the space of infinitely differentiable functions on $\BR$ with compact support.
This representation of CCR's is called the {\it Schr\"odinger representation of the CCR}
(or the {\it Schr\"odinger system} \cite{pu67}) {\it with $d$ degrees of freedom}.

By an application of  (\ref{UR}), one obtains
the {\it position-momentum 
uncertainty  relations}  
\begin{equation}
(\Delta P_j)_{\psi}( \Delta Q_j)_{\psi} \geq \half,\quad j=1,\ldots,d\label{UR1}
\end{equation}
for all $\psi\in \D(P_jQ_j)\cap \D(Q_jP_j)$ with $\|\psi\|=1$, basic inequalities in quantum
mechanics which show a big difference between quantum mechanics and classical mechanics.\footnote{
Inequality (\ref{UR1}) holds also for all $\psi \in \D(P_j)\cap\D(Q_j)$ with $\|\psi\|=1$.}

The Schr\"odinger representation $ \{P_j,  Q_{k}|j, k=1,\ldots, d\}$ 
is an irreducible  Weyl representation (\cite[Theorem 4.5.1]{pu67}; \cite[Theorem 3.12]{ara06}). 
Conversely it is known as the von Neumann uniqueness theorem (e.g., 
\cite[Theorem 4.11.1]{pu67}) that, 
if $\hhh$ is separable and $ \{A_j,  B_{k}|j, k=1,\ldots,d\}$ is an irreducible  Weyl representation
of the CCR's with $d$ degrees  of freedom, 
then 
$$
\hhh\cong \LR, \quad 
A_j\cong P_j, \quad B_j\cong Q_j, \quad j=1,...,d. $$
Here 
$\cong$ denotes  a unitary equivalence.

Usually models of quantum mechanics in $\BR$  are  constructed from   the Sch\"odinger
representation of the CCR's with $d$ degrees of freedom. In this case, 
physical quantities, which are required to be represented  by self-adjoint operators on $\LR$, are made from $P_j$ and $Q_j$, $j=1,\ldots,d$.
Among others, the  Hamiltonian of a model, which describes the total energy of the quantum system under consideration, is important. 
The classical Hamiltonian of a non-relativistic particle of mass $m$ in a potential $V:\BR \to \RR$ is
given by $ H_{\rm cl}(p,x)=p^2/2m+V(x),\, (p,x)\in \BR\times \BR$.
Then the corresponding quantum Hamiltonian
is given by
the Schr\"odinger operator (or the Schr\"odinger Hamiltonian)
$$
\HS:=H_{\rm cl}(P, Q):=\frac 1{2m} \sum_{j=1}^d P_j^2+V (Q)=-\frac 1{2m}\Delta+V(Q) 
$$
on 
$\LR$,  where  
$V (Q)$ is defined by the functional calculus
using the joint   spectral measure  of $Q_1, \cdots, Q_d$ (note that $(Q_1,\ldots,Q_d)$
is a set of strongly commuting self-adjoint operators\footnote{
A set $\{A_1,\ldots,A_n\}$ of self-adjoint operators on a Hilbert space
is said to be {\it strongly commuting} if the spectral measure $E_{A_j}$ of $A_j$
commutes with $E_{A_k}$ for all $j,k=1,\ldots,n, j\not=k$ (i.e.,
for all Borel sets $J,K\subset \RR$, $E_{A_j}(J)E_{A_k}(K)=
E_{A_k}(K)E_{A_j}(J)$).}) and $\Delta:=\sum_{j=1}^dD_j^2$ is the  $d$-dimensional
generalized Laplacian. It is shown in fact  that $V(Q)$ is the multiplication operator by
the function $V$ . Hence one simply denotes $V(Q)$ by $V$. Thus
\begin{equation}
\HS=H_0+V, \label{HV}
\end{equation}
where 
\begin{equation}
H_0:=-\frac 1{2m}\Delta.\label{H0}
\end{equation}

In general, according to an axiom of quantum mechanics due to von Neumann, 
the time evolution of the quantum system whose Hamiltonian is given by a self-adjoint operator $H$ on
 a Hilbert space $\hhh$ is described by the unitary operator $e^{-itH}$ with time parameter 
$t\in\RR$ in such a way that, if $\phi\in {\hhh}$ is a state vector at $t=0$, then
the state vector at time $t$ is given by   $\phi_t=e^{-itH}\phi$, provided that  no measurement is made for
the quantum system under consideration  in the time
interval $[0,t]$. 
If $\phi\in \D(H)$, then
$\phi_t$ is strongly differentiable in $t$, $\phi_t\in D(H)$ for all $t\in\RR$,  and
obeys the abstract Schr\"odinger equation
$$
i\frac{d \phi_t}{dt}= H\phi_t.
$$ 
Here  time $t$ is usually treated as a parameter,   not as an operator.
It is the external time mentioned in Subsection \ref{subsec-time}.    
In relativistic classical mechanics, the energy variable  is regarded as the variable canonically conjugate to 
the time variable
as so is the momentum variable to the position variable and  this may be extended to non-relativistic classical mechanics
as a limit of relativistic one.
From this point of view (or in view of the  time-energy uncertainty relation
proposed by Heisenberg), one may infer that a quantum Hamiltonian $H$
may have a symmetric  operator $T$ corresponding to time, satisfying CCR
\begin{equation}
[H,T]=-i\one\label{CCR2}
\end{equation}
on a non-zero subspace $\MSD_{H,T}$ included in $\D(HT)\cap \D(TH)$.
Such an operator $T$ is called  a {\it time operator} of  $H$ (some authors
use the form $[H,T]=i\one$ instead of (\ref{CCR2}), but this is not essential, just a convention).
From a purely mathematical point of view (apart from the context of quantum physics), this definition
applies to  any pair $(H,T)$ of a self-adjoint operator $H$ and a symmetric operator $T$
obeying (\ref{CCR2}) on a non-zero subspace included in $\D(HT)\cap \D(TH)$.

\begin{remark}\label{rem-top}{\rm It is obvious that, if $T$ is a time operator of $H$, then,
 for all $\alpha\in \RR\setminus\{0\}$, $\alpha^{-1}T$ is a time operator of $\alpha H$.}
\end{remark} 

The uncertainty relation
\begin{equation}
(\Delta H)_{\psi}(\Delta T)_{\psi}\geq \frac 12,\quad \psi\in \MSD_{T,H},\,\|\psi\|=1\label{TE-ur}
\end{equation}
implied by (\ref{CCR2}) may be interpreted as a form of  {\it time-energy
uncertainty relation}. The time operator $T$ is physical
in the sense that it gives a lower bound for the uncertainty $(\Delta H)_{\psi}$
of $H$ with respect to the state $\psi\in \MSD_{T,H}$.

In the physics literature,  formal (heuristic) constructions of  \lq\lq{time operators}"  have been done
for special classes of  Schr\"odiner Hamiltonians (e.g., \cite{ab61,fuj80,fwy80,gys81,bau83}).   
But, since  the theory of CCR's with dense CCR-domains involves unbounded operators as remarked above,
formal manipulations are questionable and results based on them remain vague and inconclusive.
In fact, mathematically rigorous considerations lead  one to
distinguish  some classes of  time operators as  recalled below. These classes correspond
to different types of  representations of CCR's (see, e.g., \cite{fug67,jm80,sch83a,sch83b,dor84}).
It should be noted that
there exist representations of CCR's which are inequivalent to Schr\"odinger ones (e.g., \cite{fug67},
\cite{sch83b}, \cite{ara98}) and, interestingly enough, some of them are connected with  characteristic 
physical phenomena such as  the so-called Aharonov-Bohm effect (see \cite{ara98} and references therein).

Mathematically rigorous  studies on time operators, including
general theories of time operators (not necessarily restricted to time operators of   Schr\"odinger operators),  have been made by some authors (e.g.,
\cite{miy01, gal02, gcb04, ara05, ara07, ara08a, ara08b, am08a,  am08b, ara09, hkm09} and references therein; see also \cite{jm80, sch83a, sch83b,dor84} for earlier studies from purely mathematical points of view).
The present paper is a continuation of those studies, in particular, concentrating
on constructions of  time operators
{\it in a generalized sense}  associated with a  class of Schr\"odinger
operators which contains the Hamiltonian of the hydrogen atom.

Let $H$ be a self-adjoint operator on $\hhh$ and  bounded from below. Then  
the von Neumann uniqueness theorem tells us that 
there exists no self-adjoint operator $T$ 
such that pair $ (H,  T)$ satisfies  the Weyl relation (\ref{Weyl}) with $d=1$,  
since  $\s (P)=\RR$ and then $H\not\cong P$, where, for a linear operator $L$, $\s(L)$ denotes the spectrum of  $L$. 
Thus, to treat such a case,  it is natural to  introduce a weaker version 
of the Weyl representation with one degree of freedom  to define a class of  
time operators.
  
\bd{wwr}\TTT{weak Weyl relation}
{\rm 
A pair $ (A,  B)$ consisting of a self-adjoint operator $A$ and symmetric operator $B$ on $\hhh$
is called a {\it weak Weyl representation} with one degree of freedom
if   
$e^{-itA} \D (B)\subset \D (B)$ for all $t\in \RR$ and the {\it weak Weyl relation} 
\begin{equation}
B e^{-itA}\psi =e^{-itA} (B+t)\psi \label{w-Weyl}
\end{equation}
holds for all $\psi\in \D (B)$ and all $t\in\RR$.  
}
\ed

Studies on this class of representations from purely mathematical points of view  have been 
done in \cite{jm80, sch83a, sch83b,dor84}.
It is easy to see  that a Weyl representation $\{A,B\}$
is a weak Weyl  representation and that
the weak Weyl  relation  (\ref{w-Weyl}) implies the CCR (\ref{CCR}) on $\D(AB)\cap \D(BA)$.
But one should note that  a weak Weyl  representation $(A,B)$
with both $A$ and $B$ being self-adjoint  is  not necessarily a  Weyl representation. 

\bd{1100}
\TTT{strong time operator}
{\rm 
A symmetric operator 
$T$ on $\hhh$ is called a {\it strong time operator} of
 a self-adjoint operator $H$ on $\hhh$ if   
$ (H, T)$ is a  weak Weyl  representation.  
}
\ed

\begin{remark}\label{rem-u-strong}{\rm (1) In relation to strong time operators,
it may be convenient to give a name to a self-adjoint operator $T$  on $\hhh$
such that $(H,T)$ is a Weyl representation  of the CCR
with one degree of freedom. We call such an operator $T$ an {\it ultra-strong time operator} of $H$.
It follows that an ultra-strong time operator is a strong time operator. But the converse is not true.
If $\hhh$ is separable, then, by the von Neumann uniqueness theorem, $(H,T)$ is unitarily equivalent to the
direct sum of the Schr\"odinger representation $(P,Q)$  with $d=1$.

(2) It is well known or easy to see that, if $(H,T)$ is a Weyl representation of the CCR with one degree of
freedom, then $\sigma(H)=\sigma(T)=\RR$ (for this fact, separability of $\hhh$ is not assumed).
Hence a semi-bounded self-adjoint operator (i.e. a self-adjoint operator which is
bounded from below or above) has no ultra-strong time operators. 
}
\end{remark}

As far as we know,  
a firm mathematical investigation of a strong time operator was initiated by \cite{miy01}, although
the name \lq\lq{strong time operator}" is not used in \cite{miy01} (it was introduced first in
\cite{ara08b} to
distinguish different classes of time operators). Further investigations and  generalizations on strong time operators
were done in 
\cite{ara05, ara07}.   See also 
\cite{am08a, am08b, hkm09,rt09}.  It is known that, if $ (H, T)$ satisfies the weak Weyl relation,  
then $\s(H)$ is purely absolutely continuous \cite{sch83a}.  
Hence, if $H$ has an eigenvalue,  then $H$ has no strong time operator.

In the context of quantum physics, in addition to time-energy uncertainty relation (\ref{TE-ur}), 
a strong time operator $T$ of a Hamiltonian $H$ may have  properties richer
than those of time operators of $H$. For example,
it controls decay rates in time $t\in\RR$  of transition probabilities $|(\phi, e^{-itH}\psi)|^2$ ($\phi,\psi\in \hhh,
\|\phi\|=\|\psi\|=1$) in the following  form \cite[Theorem 8.5]{ara05}: for each natural number 
$n\in \mathbb{N}$ and  all unit vectors $\phi,\psi\in \D(T^n)$, there exists a constant $d_n^T(\phi,\psi)\geq 0$ such that,
for all $t\in \RR\setminus\{0\}$,  
$$
|(\phi,e^{-itH}\psi)|^2\leq \frac{d_n^T(\phi,\psi)^2}{|t|^{2n}}.
$$
This shows a very interesting correspondence between  decay rates in time of
transition probabilities and regularities of state vectors $\phi, \psi$.\footnote{
Here we mean by \lq\lq{regularity}"  of a vector $\psi$ the number $n$ such that $\psi\in \D(T^n)$. 
} It tells us also the importance of domains of time operators.

In \cite{gal02,am08b}, a time operator of  a self-adjoint operator whose spectrum is  purely discrete
with a growth condition is constructed. In \cite{ara09},  necessary and sufficient conditions for a self-adjoint
operator with purely discrete spectrum to have a time operator were given.
From these investigations, it is suggested that the concept of time operator
should be weakened for a self-adjoint operator (a Hamiltonian in the context of quantum mechanics) 
whose spectrum is not purely absolutely continuous and whose discrete spectrum does not satisfy
conditions formulated in \cite{ara09}.
One of weaker versions of time operator is defined as follows:

\bd{weakdef}\TTT{weak time operator}
{\rm A symmetric operator $T$ on $\hhh$  is called 
a {\it weak time operator} of  
a self-adjoint operator $H$ on $\hhh$  if there exists a non-zero subspace $\MSD_{\rm w}
\subset \D(T)\cap\D(H)$ such that the {\it weak CCR} on $\MSD_{\rm w}$ holds:
\begin{equation}
(H\phi,  T\psi)- (T\phi,  H\psi)=-i (\phi, \psi),\quad \phi,\psi\in \MSD_{\rm w}. \label{w-CCR}
\end{equation}
We call $\MSD_{\rm w}$ a {\it weak-CCR domain} for the pair $(H,T)$.
}
\ed

It is obvious that a time operator $T$ of  $H$ is a weak time operator
of  $H$ with $\MSD_{\rm w}=\MSD_{H,T}$.
We remark that (\ref{w-CCR}) implies the time-energy uncertainty relation (\ref{TE-ur})
with $\psi\in\MSD_{\rm w}$ ($\|\psi\|=1$).

One should keep in mind the following fact:

\begin{proposition}\label{prop-w-time} Let $T$ be a weak time operator of a self-adjoint operator $H$ and
$\MSD_{\rm w}$ be a weak-CCR domain for $(H,T)$. Then $H$ has no eigenvectors in $\MSD_{\rm w}$.
\end{proposition}

\proof Let $H\psi=E\psi$ with $\psi\in \MSD_{\rm w}$ and $E\in \RR$.
Taking $\phi $ in (\ref{w-CCR}) to be $\psi$,  we see that the left hand side is equal to  0. Hence
$\|\psi\|^2=0$, implying $\psi=0$.\qed

\begin{remark}{\rm Unfortunately we do not know whether  or not there exists a weak time operator which cannot
be a time operator. We leave this problem  for future study.
}
\end{remark}
 
\subsection{Ultra-weak time operator}\label{subsec-uwt}

Proposition \ref{prop-w-time} implies that, if 
a self-adjoint operator $H$ with an eigenvalue $E$  has a weak time operator,
then all the eigenvectors of $H$ with eigenvalue $E$ are  out of any weak-CCR domain for $(H,T)$.
On the other hand, $H$ may have a complete set of  eigenvectors so that
the subspace  algebraically spanned by the eigenvectors of $H$ is dense in $\hhh$. 
This suggests that
such a  self-adjoint operator may have tendency not to have a weak time operator.
Taking into account this possibility and in the spirit of seeking   ideas as general as possible, 
we generalize the concept of   weak time operator:

\begin{definition}[ultra-weak time operator]\label{def-uwt}{\rm Let $H$ be a self-adjoint operator on $\hhh$ and 
$\MSD_1$ and $\MSD_2$ be  non-zero subspaces of $\hhh$.
A sesquilinear form $\mft:\MSD_1 \times \MSD_2\to\CC$ ($\MSD_1\times\MSD_2\ni (\phi,\psi)\mapsto \mft[\phi,\psi]\in\CC$)  
with domain $\D(\mft)=\MSD_1\times \MSD_2$ 
($\mft[\phi,\psi]$ is antilinear in $\phi$ and  linear in $\psi$)
is called  {\it  an ultra-weak time operator} of  $H$ if
there exist non-zero subspaces $\MSD$ and 
$\ms E$ of $\MSD_1\cap\MSD_2$ such that
the following (i)--(iii) hold:
\begin{list}{}{}
\item[(i)] $\ms E \subset D(H)\cap \MSD$.
\item[(ii)] (symmetry on $\MSD$) $\mft[\phi,\psi]^*=\mft[\psi,\phi],\, \phi,\psi\in \MSD$. 
\item[(iii)] (ultra-weak CCR)  $H\ms E\subset \MSD_1$ and, for all $\psi, \phi\in \ms E$, 
\begin{equation}
\mft[H\phi,\psi] -\mft[H\psi,\phi]^*=-i(\phi,\psi)\label{TH}
\end{equation}
\end{list}
We call $\ms E$  an {\it ultra-weak CCR-domain} for $(H,\mft)$
and $\MSD$ a {\it symmetric domain} of $\mft$.  
}
\end{definition}

\begin{remark}\label{rem-uwt}{\rm 
(1) As far as we know, the concept \lq\lq{ultra-weak time operator}" introduced here is new.

(2) Although there may be no operators associated with the sesquilinear form $\mft$
in the above definition,
we use, by abuse of word, \lq\lq{ultra-weak time operator}" to indicate
that it is a  concept weaker than that of  weak time operator as shown below.

 Let $T$  be  a weak time operator of  $H$ with a weak CCR-domain $\MSD_{\rm w}$.
Then one can define a sesquilinear form $\mft_T:\hhh \times \D(T)
\to \CC$ by
$$
\mft_T[\phi,\psi]:=(\phi,T\psi),\quad \phi\in\hhh, \psi\in \D(T).
$$
Then it is easy to see that
$\mft_T[ \phi,\psi]^*=\mft_T[ \psi,\phi],\,\psi,\phi\in \D(T)$ and, for all $\phi,\psi\in \MSD_{\rm w}$,
 $\mft_T[ H\phi,\psi]-\mft_T[ H\psi,\phi]^*=-i(\psi,\phi)$. Hence $\mft_T$ is an ultra-weak  time operator
 of  $H$ with $\MSD_{\rm w}$ being an ultra-weak CCR-domain and $\D(T)$ a symmetry domain. Therefore the concept of ultra-weak time operator is  a generalization of  weak time
 operator.

(3) If $H\psi\in \MSD$ in (\ref{TH}), then, by the  symmetry of $\mft[\cdot,\cdot]$ on $\MSD$, 
(\ref{TH}) takes the following form:
$$
\mft[ H\phi,\psi] -\mft[ \phi,H\psi]=-i(\phi,\psi)
$$
}
\end{remark}

For a sesquilinear form $\mft:\MSD_1\times \MSD_2 \to \CC$ and  a constant $a\in\RR$, we define a sesquilinear form $\mft-a:\MSD_1\times \MSD_2\to\CC$
by
$$
(\mft-a)[\phi,\psi]:=\mft[\phi,\psi]-a(\phi,\psi),\quad \phi\in \MSD_1,\psi\in \MSD_2.
$$

In the case of the pair $(H,\mft)$ 
in Definition \ref{def-uwt},   the uncertainty relation (\ref{UR})  associated with CCR 
is generalized  as follows:

\begin{proposition}[uncertainty relation for $(H,\mft)$] \label{U1}
Assume that $H$ has an ultra-weak time operator $\mft$ as 
in Definition \ref{def-uwt}.
Then, for all $a,b\in\RR$ and a unit vector $\psi\in\ms E$,
\begin{equation}
\left|(\mft-a)[ (H-b)\psi,\psi]\right|\geq \frac 12.\label{TH1}
\end{equation}
\end{proposition}

\proof Using (\ref{TH}), we have $\Im \left\{(\mft-a)[ (H-b)\psi,\psi]\right\}
=-\frac 1{2}$.
Since $|z|\geq |\Im z|$ for all $z\in\CC$, (\ref{TH1}) follows.
\qed

In summary, we have seen that there exist five classes of time operators
with inclusion relation (\ref{hierarchy}).

\subsection{Outline of the present paper}

Having introduced the new concept \lq\lq{ultra-weak time operator},
we now outline the contents of the present paper.
In Section \ref{sec3}, we review   an abstract theory of time operators
and give new additional   results.
Among others, we prove an existence theorem
on a strong time operator of an absolutely continuous self-adjoint operator
(Theorem \ref{thm-ac-st}).
Sections \ref{sec4} is devoted to showing the existence of  time operators of self-adjoint operators with purely discrete spectra.
This includes  an extension of existence theorems on time operators
in \cite{gal02,am08b}.
In Section \ref{sec5}, we introduce a class $S_1(\hhh)$ of self-adjoint operators on $\hhh$ (see Definition \ref{def-S1}) such that
each element of $S_1(\hhh)$ has an ultra-weak time operator with a dense ultra-weak CCR-domain (Theorem \ref{thm-uwt1}).
Moreover, for a class of Borel measurable functions $f:\RR\to\RR$, we formulate
sufficient conditions for $f(H)$ to have an ultra-weak time operator (Corollary \ref{main3}).
In Section \ref{sec6}, we discuss applications of the abstract results  to
the Schr\"odinger operator $\HS$.
We find classes of potentials $V$ for which
$\HS$ has an ultra-weak time operator with a dense ultra-weak CCR-domain (Theorem \ref{thm-uwt2}). 
Also we show that the Hamiltonian of the  hydrogen atom
(i.e. the case where $V(x)=-\gamma/|x|, \, x\in \RR^3\setminus\{0\}$ with a constant
$\gamma>0$) has an ultra-weak time operator with a dense ultra-weak CCR-domain (Example \ref{ex-hyd2}).
In the last section, for a class of $f$, an existence theorem on an ultra-weak time operator of $f(\HS)$
is proved (Theorem \ref{thm-uwt-HS}) and some examples are given.

\section{Abstract Theory of Time Operators--Review with Additional Results}
\label{sec3}

\subsection{A general structure of time operators}

We first note an elementary fact:

\begin{proposition}\label{prop-uni} Let $H$ be a self-adjoint operator on a Hilbert space $\hhh$  and $T$ be a time
operator of $H$ with a CCR-domain $\MSD$ for $(H,T)$.
Let $H'$ be a self-adjoint operator on a Hilbert space $\hhh'$ such that $UHU^{-1}=H'$ for a unitary
operator $U:\hhh\to\hhh'$. Then $T':=UTU^{-1}$
is a time operator of $H'$ with a CCR-domain $U\MSD$ for $(H',T')$.
\end{proposition}

\proof An easy exercise.
\qed

In what follows, $H$ denotes a self-adjoint operator on a complex Hilbert space $\hhh$.
As is well known (e.g., \cite[\S 10.1]{ka76}, \cite[Theorem VII.24]{rs1}), $\hhh$ has the orthogonal decomposition
\begin{equation}
\hhh=\hhh_{\rm ac}(H)\oplus\hhh_{\rm sc}(H)\oplus \hhh_{\rm p}(H),
\end{equation}
where $\hhh_{\rm ac}(H)$ (resp. $\hhh_{\rm sc}(H)$, $\hhh_{\rm p}(H)$) is
the subspace of absolute continuity (resp. of singular continuity, of  discontinuity) with respect to $H$,
and $H$ is reduced by each subspace $\hhh_{\#}(H)$ ($\#=$ ac, sc, p).
We denote the reduced part of $H$ to $\hhh_{\#}(H)$ by $H_{\#}$ and set
$$
\sigma_{\rm ac}(H):=\sigma(H_{\rm ac}),\quad \sigma_{\rm sc}(H):=\sigma(H_{\rm sc}),
$$
which are  called the absolutely  continuous spectrum and the singular continuous spectrum  of $H$ respectively.
We denote by $\sigma_{\rm p}(H)$ the set of all eigenvalues of $H$. We remark
that $\sigma(H_{\rm p})=\overline{\sigma_{\rm p}(H)}$, the closure of $\sigma_{\rm p}(H)$.
We have
\begin{equation}
H=H_{\rm ac}\oplus H_{\rm sc}\oplus H_{\rm p}\label{H-decomp}
\end{equation}
and
$$
\sigma(H)=\sigma_{\rm ac}(H)\cup\sigma_{\rm sc}(H)\cup \overline{\sigma_{\rm p}(H)}.
$$

An eigenvalue of $H$ is called a discrete eigenvalue of $H$ if it is an isolated eigenvalue of $H$ with a finite
multiplicity. The set $\sigma_{\rm disc}(H)$ of all the discrete eigenvalues of $H$ is called the
discrete spectrum of $H$. 

The following proposition shows that the problem of constructing  time operators
of $H$ is reduced to that of constructing time operators of each $H_{\#}$.

\begin{proposition} Suppose that each $H_{\#}$ has a time operator $T_{\#}$ with a CCR-domain $\MSD_{\#}$. Then
the direct sum
$$
T:=T_{\rm ac}\oplus T_{\rm sc}(H)\oplus T_{\rm p}
$$
is a time operator of $H$ with a CCR-domain $\MSD_{\rm ac}\oplus \MSD_{\rm sc}\oplus \MSD_{\rm p}$.
\end{proposition}

\proof Since the direct sum of symmetric operators is again a symmetric operator in general,
it follows that  $T$ is symmetric.
By the assumption, we have for all $\psi_{\#}\in \MSD_{\#}$
$$
[H_{\#},T_{\#}]\psi_{\#}=-i\psi_{\#}.
$$
Let $\psi=(\psi_{\rm ac},\psi_{\rm sc},\psi_{\rm p})\in \MSD_{\rm ac}\oplus \MSD_{\rm sc}\oplus \MSD_{\rm p}$.
Then, by (\ref{H-decomp}), $\psi \in D(HT)\cap D(TH)$ and
$$
[H,T]\psi=([H_{\rm ac},T_{\rm ac}]\psi_{\rm ac},[H_{\rm sc},T_{\rm sc}]\psi_{\rm sc},
[H_{\rm p},T_{\rm p}]\psi_{\rm p})=-i\psi.
$$
Hence $T$ is a time operator of $H$ with a CCR-domain $\MSD_{\rm ac}\oplus \MSD_{\rm sc}\oplus \MSD_{\rm p}$.
\qed

\subsection{Strong time operators}

\subsubsection{A summary of known results and additional results}

We  summarize some basic facts on strong time operators of $H$.

\begin{proposition}\label{prop-st} A symmetric operator $T$ is a strong time operator of $H$ if and only if
operator equality $e^{itH}Te^{-itH}=T+t$ holds for all $t\in\RR$.
\end{proposition}

\proof See \cite[Proposition 2.1]{ara05}.
\qed

Note that the operator equality given in this proposition
implies that, for all $t\in \RR$, $e^{-itH}\D(T)=\D(T)$.

\begin{proposition}\label{prop-st-uni} Let $T$ be a strong time operator of $H$
and $H'$ be a self-adjoint operator on a Hilbert space $\hhh'$ such that,
for a unitary operator $U:\hhh\to\hhh'$, $UHU^{-1}=H'$.
Then $T':=UTU^{-1}$ is a strong time operator of $H'$.
\end{proposition}

\proof By the functional calculus,  for all $t\in \RR$, $e^{itH'}=Ue^{itH}U^{-1}$. By this fact and
Proposition \ref{prop-st},  
we have
$$
e^{itH'}T'e^{-itH'}=Ue^{itH}Te^{-itH}U^{-1}=U(T+t)U^{-1}=T'+t.
$$
Hence $T'$ is a strong time operator of $H'$.
\qed

\begin{proposition}[\cite{ara05}]
\label{time}
Suppose that $H$ has a strong time operator 
$T$.  
Then:
\begin{list}{}{}
\item{(1)} The closure $\overline{T}$ of $T$  is also a strong 
time  operator of $H$. 
\item{(2)} If $H$ is semi-bounded, then $T$ is not essentially self-adjoint.  
\item{(3)} The operator $H$ is absolutely continuous.  
\end{list}
\ep

\begin{proposition}\label{prop-st1} Let $T_1,\ldots, T_n$ ($n\geq 2$) be strong time operators of $H$.
\begin{list}{}{}
\item{(1)} Let $S:=\sum_{k=1}^na_kT_k$ with $a_k\in\RR$ ($k=1,\ldots,n$) satisfying \linebreak $\sum_{k=1}^na_k=1$.
Then, for all $t\in \RR$, operator equality
\begin{equation}
e^{itH}Se^{-itH}=S+t\label{S}
\end{equation}
holds. In particular, if $\cap_{k=1}^nD(T_k)$ is dense, then $S$ is a strong time operator of $H$.
\item{(2)} For any pair $(k,\ell)$ with $k\not=\ell$ ($k,\ell=1,\ldots,n$), 
$(T_k-T_{\ell})e^{itH}\psi=e^{itH}(T_k-T_{\ell})\psi$ for all $t\in \RR$ and $\psi\in D(T_k)\cap D(T_{\ell})$.
\end{list}
\end{proposition}

\proof (1) By Proposition \ref{prop-st}, we have operator equalities 
\begin{equation}
e^{itH}T_ke^{-itH}=T_k+t, \quad 
t\in \RR, k=1,\ldots,n. \label{Tk}
\end{equation}
Since $e^{itH}Se^{-itH}=\sum_{k=1}^ne^{itH}a_kT_ke^{-itH}$ (operator equality),
(\ref{Tk}) implies (\ref{S}).
If  $\cap_{k=1}^d\D(T_k)$ is dense, then $S$ is a symmetric
operator and hence it is a strong time operator of $H$.

(2) This easily follows from (\ref{Tk}).
\qed

Proposition \ref{prop-st1}-(1) shows that any real convex combination $S$ of  strong time operators of $H$
such that  $\D(S)$ is dense is a strong time operator of $H$.

Let $\{H_1, \ldots, H_n\}$ be a set of strongly commuting self-adjoint operators
on $\hhh$. Then 
$\sum_{j=1}^nH_j$ 
is essentially self-adjoint
and, for all $t\in \RR$,
\begin{equation}
e^{it\overline{\sum_{j=1}^nH_j}}=\prod_{j=1}^ne^{itH_j},\label{expH}
\end{equation}
where the order of the product of
$e^{itH_1},\ldots, e^{itH_n}$ on the right hand side is arbitrary (this is due to the commutativity of
$e^{itH_j}$ and $e^{itH_k}$ ($j,k=1,\ldots,n$) which follows the strong commutativity of $\{H_1,\ldots,H_n\}$).

\begin{proposition}\label{prop-st2} Let $\{H_1,\ldots,H_n\}$ be as above  and assume that, for some $j$,
$H_j$ has a strong time operator $T_j$ such that
$e^{itH_k}T_je^{-itH_k}=T_j$ for all $k\not=j$. Then $T_j$ is a strong time operator of $\overline{\sum_{j=1}^nH_j}$.
\end{proposition}

\proof By the present assumption and Proposition \ref{prop-st}, we have 
operator equality $e^{itH_j}T_je^{-itH_j}=T_j+t$ for all $t\in \RR$.
Hence, by (\ref{expH}) and the commutativity of the operators $e^{itH_k}, \, k=1,\ldots,n$,
we have 
$$
e^{it\overline{\sum_{j=1}^nH_j}}T_je^{-it\overline{\sum_{j=1}^nH_j}}=
\left(\prod_{k\not=j}e^{itH_k}\right)(T_j+t)\left(\prod_{k\not=j}e^{-itH_k}\right)
=T_j+t.
$$
Thus the desired result follows.
\qed

Proposition \ref{prop-st2} may be useful
to find  strong time operators of a self-adjoint operator which is given by the closure of the sum
of strongly commuting self-adjoint operators. 

A variant of Proposition \ref{prop-st2} is formulated as follows.
Let  $\{A_1,\ldots, A_n\}$ be a set of strongly commuting self-adjoint operators
on $\hhh$ such that each $A_j$ is injective. Suppose that each $A_j$  has a strong time operator $B_j$
such that, for all $j=1,\ldots,n$, $\D({B}_jA_j^{-1})\cap \D(A_j^{-1}{B}_j)$ 
is dense and, for all $t\in \RR$,  
$e^{itA_k}B_je^{itA_k}=B_j, \, k\not=j, k=1,\ldots,n$.
By the strong commutativity of $\{A_1,\ldots,A_n\}$,
the operator
$$
H_A:=\sum_{j=1}^nA_j^2
$$
is a non-negative self-adjoint operator.
For each $j=1,\ldots,n$, the operator
$$
T_j:=\frac 14 \left(\overline{B}_jA_j^{-1}+A_j^{-1}\overline{B}_j\right)
$$
is symmetric.

\begin{proposition}[\cite{ara05}]\label{prop-ara05}
For each $j=1,\ldots,n$, $T_j$ is a strong time operator of $H_A$.
\end{proposition}

A general scheme to construct strong time operators for a given pair $(H,T)$ of  a weak Weyl representation
is described in \cite[\S 10]{ara05}. A generalization of this scheme is given as follows.
By the functional calculus, for any real-valued continuous function $f$ on $\RR$,
$f (H)$ is a self-adjoint operator on $\hhh$.
Then a natural question is: does $f(H)$ has a strong time operator ?
A heuristic argument to answer the question is as follows. Let $f\in C^1(\RR)$ and denote the derivative of $f$ by $f'$. We have
$[T, H]=+i\one $,  which intuitively implies that $T=+id/dH$.  
Hence we may formally see that 
$[T,  f (H)]=if' (H)$(in \cite[Theorem 6.2]{ara05}, this is justified for
all $f\in C^1(\RR)$ such that $f$ and $f'$ are bounded), and then 
 $Te^{-itf (H)}=e^{-itf (H)} (T+tf' (H))$ holds.  
Multiplying 
$f' (H) ^{-1}$ on the both sides, we may have 
$T f' (H) ^{-1} e^{-itf (H)}=
e^{-itf (H)} (Tf' (H) ^{-1}+t)$,  and, by symmetrizing $T f' (H) ^{-1}$,
we expect that 
$\half (T f' (H) ^{-1}+f' (H) ^{-1}T)$ is 
 a strong time operator of  $f (H)$. 
Actually this result  is  justified under some conditions:
 
\begin{proposition}[{\cite[Theorem 1.9]{hkm09}}]\label{prop-hkm}
\label{hkm09}
Let $K$ be a closed null subset of $\RR$ with respect to the Lebesgue measure.
Assume that  $f\in C^2 (\RR\setminus K)$ and 
$L:=\{\lambda\in \RR\setminus K| f' (\lambda)=0\}$ is a null set with respect to the Lebesgue measure. 
Suppose that $H$ has a strong time operator $T_H$ which is closed and let
 $$
D:=\{g(H) \D (T_H)| g\in C_0^\infty (\RR\setminus L\cup K)\}.
$$  
 Then 
$$
T_{f (H)}:=\half\overline{(T_Hf' (H)^{-1}+f' (H)^{-1}T_H) \lceil D}
$$ 
is a strong time operator of  $f (H)$,
where, for a linear operator $L$ and a subspace $\MSD\subset \D(L)$,
$L\lceil \MSD$ denotes the restriction of $L$ to $\MSD$.
\ep

\begin{example}[Aharonov-Bohm time operator]\label{ex-AB}{\rm Let $m>0$ be a constant.
Then it is obvious that $\sqrt{2m}Q_j$ is a strong time operator of  $P_j/\sqrt{2m}$
in the Hilbert space  $\hhh=L^2(\BR)$.
Consider the function $f(\lambda)=\lambda^2,\,\lambda\in \RR$. Then $f'(\lambda)=2\lambda$.
Hence $\{\lambda\in\RR|f'(\lambda)=0\}=\{0\}$.
Therefore the subspace $D$ in Proposition \ref{prop-hkm} takes the form
$D_{{\rm AB},j}:={\rm L.H.}\{g(P_j)\D(Q_j)|g\in C_0^{\infty}(\RR\setminus \{0\})\}$.
Hence, letting
$$
T_{{\rm AB},j}:=\frac m2\left(Q_jP_j^{-1}+P_j^{-1}Q_j\right),
$$
the operator 
$$
\widetilde T_{{\rm AB},j}:=\overline{T_{{\rm AB},j}\lceil D_{{\rm AB},j}}
$$
is a strong time operator of $P_j^2/2m$. 
Since $(P_1,\ldots,P_d)$ is a set of strongly commuting self-adjoint operators,
it follows from  Proposition \ref{prop-ara05} that
$\widetilde T_{{\rm AB},j}$ is a strong time operator of $H_0$.

There is another domain on which $T_{{\rm AB},j}$ becomes  a strong time operator of $H_0$ \cite{ara07}.
Let 
$$\Omega_j:=\{k\in \BR| k_j\not=0\}, \quad D_{{\rm AB},j}':=\{f\in L^2(\BR)|\hat f\in C_0^{\infty}(\Omega_j)\},$$
where $\hat f$ is the $L^2$-Fourier transform of $f$. 
Then $D_{{\rm AB},j}'$ is dense. Moreover, by using the Fourier analysis, it is shown that  
the operators $Q_j,P_j^{-1},  e^{itP_j^2/2m}$ and
$e^{itH_0}$ ($\forall t\in\RR$) leave $D_{{\rm AB},j}'$ invariant and, for all $t\in \RR$, 
$e^{itH_0}T_{{\rm AB},j}e^{-itH_0}=T_{{\rm AB},j}+t$ on $D_{{\rm AB},j}'$.
Hence 
$$
T_{{\rm AB},j}':=T_{{\rm AB},j}\lceil D_{{\rm AB},j}'
$$ is a strong time operator of
$H_0$. We note that $\D(Q_j)\supset D_{{\rm AB},j}'$. Hence, for each $g\in C_0^{\infty}(\RR\setminus \{0\})$,
$g(P_j)\D(Q_j)\supset g(P_j)D_{{\rm AB},j}'$. For any $g\in C_0^{\infty}(\RR\setminus\{0\})$ 
such that $\hat g(k_j)>0,\forall k_j\in \RR$, $g(P_j)D_{{\rm AB},j}'=D_{{\rm AB},j}'$. It is not so difficult to show 
that such a function $g$ exists.
Therefore $D_{{\rm AB},j}\supset D_{{\rm AB},j}'$ in fact.
A time operator of $H_0$ obtained  as a restriction  of $T_{{\rm AB},j}$ to a subspace or its closure
is called an {\it Aharonov-Bohm time operator} \cite{ab61,miy01}. 
}
\end{example}

\begin{example}{\rm As a generalization of Aharonov-Bohm time operators,
one can construct strong time operators of a self-adjoint operator $H$ of the form $H=F(P)$ with $F\in C^1(\BR)$,
which includes the free relativistic Schr\"odinger Hamiltonian $(-\Delta +m^2)^{1/2}$ ($m>0$)
and its fractional version $(-\Delta+m^2)^{\alpha}$ ($\alpha>0$).
This approach can be  applied also to constructions of strong time operators of Dirac type operators \cite{th92}. 
See \cite[\S 11]{ara05}) for the details. }
\end{example}

 \subsubsection{Existence of a strong time operator for a class of absolutely continuous self-adjoint
 operators}
\label{4.1}
As already mentioned, a self-adjoint operator which has a strong time operator
is absolutely continuous. Then a natural question is: does an absolutely continuous self-adjoint
operator have a strong time operator ?
To our best knowledge, this question has not been answered
in an abstract framework.
In what follows, we give a partial affirmative answer to the question.

We recall an important concept.
For a linear operator $A$ on a Hilbert space  $\hhh$,
a non-zero vector $\phi\in \cap_{n=1}^{\infty}\D(A^n)$
is called a {\it cyclic vector} for $A$ if 
$${\rm L.H.}\{A^n\phi|n\in \{0\}\cup\mathbb{N}\}$$
is dense in $\hhh$, where, for a subset $\MSD$ of $\hhh$, ${\rm L.H.}\MSD$ denotes
the algebraic linear hull of vectors in $\MSD$. 

We denote by $E_H$ the spectral measure of $H$. 
For a non-zero vector $\psi\in\hhh$, a measure $\mu_{\psi}$ on $\RR$ is defined by
$$
\mu_{\psi}(B):=\|E_H(B)\psi\|^2,\quad B\in \bm{B},
$$
where $\bm{B}$ is the family of  Borel sets of $\RR$.
We define a function $X$ on $\RR$ by
$$
X(\lambda):=\lambda,\quad \lambda\in\RR.
$$
We note the following fact:

\begin{lemma}\label{lem-c1} Assume that $\hhh$ is  separable.  Suppose that $H$  has a cyclic vector $\phi$.
Then there exists a unitary operator $U$ from $\hhh$ to $L^2(\RR,d\mu_{\phi})$
such that $U\phi=1$ and $UHU^{-1}=M_X$, the multiplication operator by the function $X$ acting in
$L^2(\RR,d\mu_{\phi})$. Moreover,
the subspace ${\rm L.H.}\{e^{itX}|t\in \RR\}$ is dense in
$L^2(\RR,d\mu_{\phi})$.
\end{lemma}

\proof The first half of the lemma  follows from  an easy extension of Lemma 1 in  \cite[\S VII.2]{rs1}
to the case of unbounded self-adjoint  operators \cite[Theorem 1.8]{ara06}.
To prove the second half of the lemma, we note that, by the cyclicity of $\phi$ for $H$,
${\rm L.H.}\{H^n\phi|n\in \{0\}\cup\mathbb{N}\}$ is dense in $\hhh$.
By the functional calculus, we have
$$
\lim_{t\to 0}(-i)^n\left(\frac{e^{itH}-1}{t}\right)^n\phi=H^n\phi.
$$
Hence it follows that ${\rm L.H.}\{e^{itH}\phi|t\in\RR\}$ is dense in $\hhh$.
By the first half of the lemma, we have $Ue^{itH}\phi=e^{itX}$. Hence ${\rm L.H.}\{e^{itX}|t\in\RR\}$
is dense in $L^2(\RR,d\mu_{\phi})$.

\qed

Let $\psi \in \hhh$. If $\mu_{\psi}$ is absolutely continuous with respect to the Lebesgue measure on $\RR$,
then we denote by $\rho_{\psi}$ the  Radon-Nykod\'ym derivative of $\mu_{\psi}$: $\rho_{\psi}\geq 0$ and
$\mu_{\psi}(B)=\int_B\rho_{\psi}(\lambda)d\lambda,\, B\in \bm{B}$.

We introduce a class  of self-adjoint operators on $\hhh$. 

\begin{definition}\label{S1}
{\rm 
We say that a self-adjoint operator $H$ on $\hhh$ is in the class $S_0(\hhh)$
if it satisfies the following (i) and (ii):
\begin{list}{}{}
\item[(i)] $H$ is absolutely continuous.
\item[(ii)] $H$ has a cyclic vector $\phi$ such that
$\rho_{\phi}$ is differentiable on  $\RR$ and
\begin{align*}
& \lim_{\lambda\to\pm\infty}\rho_{\phi}(\lambda)=0,\quad
\int_{\rho(\lambda)>0}\frac{\rho_{\phi}'(\lambda)^2}{\rho_{\phi}(\lambda)}d\lambda<\infty.
\end{align*}
\end{list}
}
\end{definition}

Let $\hhh$ be separable and $H\in S_0(\hhh)$ with a cyclic vector $\phi$
satisfying the above (ii) and
$$
W_{\phi}(\lambda):=\left\{
\begin{array}{ll}
\displaystyle \frac{\rho_{\phi}'(\lambda)}{\rho_{\phi}(\lambda)} & \ \mbox{\rm for $ \rho_{\phi}(\lambda)>0 $ } \\
0 & \ \mbox{\rm for $\rho_{\phi}(\lambda)=0 $}
\end{array}
\right..
$$
Then we define  an  operator $Y$ on $L^2(\RR,d\mu_{\phi})$
 as follows:
\begin{align*}
&D(Y):={\rm L.H.}\{e^{itX}|t\in \RR\},\quad Y:=i\frac{d}{d\lambda}+\frac i2W_{\phi}.
\end{align*}

\begin{lemma} The operator $Y$ is a symmetric operator.
\end{lemma}

\proof By Lemma \ref{lem-c1}, $D(Y)$ is dense in $L^2(\RR,d\mu_{\phi})$.
Using (ii) and integration by parts, we see that, for all $f,g\in D(Y)$,
$(f,Yg)_{L^2(\RR,d\mu_{\phi})}=(Yf,g)_{L^2(\RR,d\mu_{\phi})}$.
Hence $Y$ is a symmetric operator.
\qed

\begin{lemma}\label{lem-Y} The operator $Y$ is a strong time operator of $M_X$.
\end{lemma}

\proof It is obvious that, for all $t\in\RR$, $e^{itM_X}D(Y)\subset D(Y)$.
Let $f(\lambda)=e^{is\lambda},\, s\in\RR, \lambda \in \RR$.
Then, using the fact that $if'(\lambda)=-sf(\lambda)$, we see that 
$$
(e^{itM_X}Ye^{-itM_X}f)(\lambda)=e^{it\lambda}\left(i\frac{d}{d\lambda}+\frac i2 W_{\phi}
\right)e^{-i(t-s)\lambda}
=tf(\lambda)+(Yf)(\lambda).
$$
Thus $Y$ is a strong time operator of $M_X$.
\qed

\begin{theorem}\label{thm-ac-st} Assume that $\hhh$ is separable and that $H\in S_0(\hhh)$. 
Then $H$ has  a strong time operator.
\end{theorem}

\proof We have $U^{-1}M_XU=H$. By Lemma \ref{lem-Y}, $Y$ is a strong time operator of $M_X$.
Hence, by an application of Proposition \ref{prop-st-uni}, $U^{-1}YU$ is a strong time operator of $H$.\qed

Thus we have found a class $S_0(\hhh)$ of self-adjoint operators on a separable Hilbert space $\hhh$
which each  have a strong time operator.

\subsection{Construction of strong time operators of a self-adjoint operator
from those of another self-adjoint operator}

We consider two self-adjoint operators  $H$ and $H'$ acting in  Hilbert spaces $\hhh$  and  $\hhh'$ respectively.
If $\hhh=\hhh'$, then $H'=H+(H'-H)$ on $\D(H)\cap \D(H')$
and hence $H'$ can be regarded as a perturbation of $H$. 

We denote by $P_{\rm ac}(H)$ the orthogonal projection onto the absolutely continuous subspace $\hhh_{\rm ac}(H)$ of $H$.
For a linear operator $A$, we denote by ${\rm Ran}(A)$ the range of $A$.

\begin{lemma}\label{lem-wave} Assume the following (A.1)--(A.3):
\begin{list}{}{}
\item{(A.1)} The wave operators
$$
W_{\pm}:=\mbox{\rm s-}\lim_{t\to\pm\infty}e^{itH'}Je^{-itH}P_{\rm ac}(H)
$$
exist, where $\mbox{\rm s-}\lim$ means strong limit and
$J:\hhh\to\hhh'$ is a bounded linear operator.
\item{(A.2)} $\lim_{t\to\pm\infty}\|Je^{-itH}P_{\rm ac}(H)\psi\|=\|P_{\rm ac}(H)\psi\|,\quad \psi\in \hhh$.
\item{(A.3)}(completeness)  ${\rm Ran}(W_{\pm})=\hhh_{\rm ac}(H')$.
\end{list}
Let $U_{\pm}:=W_{\pm}\lceil \hhh_{\rm ac}(H)$.
Then  $U_{\pm}$ are unitary operators from $\hhh_{\rm ac}(H)$ to $\hhh_{\rm ac}(H')$
such that
$$
H'_{\rm ac}=U_{\pm}H_{\rm ac}U_{\pm}^{-1}.
$$
\end{lemma}

\proof See textbooks of quantum scattering theory (e.g., \cite{ku79,rs3}).
\qed

\begin{theorem}\label{thm-st-uni1} Assume (A.1)--(A.3) in Lemma \ref{lem-wave}.
Suppose that $H_{\rm ac}$ has a strong time operator $T$.
Then $T'_{\pm}:=U_{\pm}TU_{\pm}^{-1}$ are strong time operators of $H_{\rm ac}'$.
\end{theorem}

\proof This follows from Lemma \ref{lem-wave} and an application of Proposition \ref{prop-st-uni}.
\qed

Theorem \ref{thm-st-uni1} can be used to construct strong time operators
of $H'$ from those of $H$.

\section{Time Operators of a Self-adjoint Operator with  Purely Discrete Spectrum }\label{sec4}

\subsection{Case (I)}
\label{weak}
%Weak time operators}

If  $\sigma_{\rm disc}(H)\not=\emptyset$,   then no strong time operator of  $H$ exists by Proposition \ref{time}-(3).  
But, even in that case, $H$ may have time operators or weak time operators.
We first recall basic results on this aspect.

\begin{proposition}[\cite{ara09, gal02}]\label{ag}
Suppose  that 
 $\s (H)=\s_{\rm disc} (H)=\{E_n\}_{n=1}^\infty$ ($E_n\not=E_m$ for $n\not=m$),  
each eigenvalue $E_n$ is simple,  
and, for some $N\geq 1$, $E_n\not=0,\,n\geq N$, 
$ \sum_{n=N}^\infty {1}/{E_n^2}<\infty$.  
Let $e_n$ be a normalized eigenvector of  $H$ with eigenvalue $E_n$: $He_{n}=E_ne_{n}$  
and 
define 
\eq{tdef}
T\phi=i\sum_{n=1}^\infty 
 \lk \sum_{m\ne n}\frac{ (e_{m},  \phi)}{E_n-E_m}\rk 
e_{n},\quad \phi\in \D(T)
\en
with   domain 
\eq{F}
\D (T):=\ms F
:={\rm L. H. }\{e_n|n\in{\mathbb N}\}, 
\en
Then $T$ is a symmetric operator and
$[H,  T]=-i\one$ holds on 
$$\ms E
:={\rm L. H.   }\{e_n-e_m|n, m\in{\mathbb N}\}.$$
Furthermore 
$\ms E$ is dense. 
\ep

This proposition shows that $T$ is a time operator of $H$ with a dense CCR-domain $\ms E$ and
hence $T$ is a weak time operator of $H$ too with a weak-CCR domain $\ms E$.
But $\D(T)=\D(T)\cap \D(H)$ cannot be  a weak-CCR domain for $(H,T)$, since
$\D(T)$ contains an eigenvector of $H$ (see Proposition \ref{prop-w-time}) (note that $\ms E$ contains
no eigenvectors of $H$).

\begin{example}[1-dimensional quantum harmonic oscillator]\label{ex-1hos}
{\rm The Hamiltonian \, of a $1$-dimensional quantum
harmonic oscillator is given by
$$
H_{\rm osc}:=-\half\Delta+\frac 12 \omega^2 x^2
$$
acting in  $L^2(\RR)$, where $\Delta$ is the $1$-dimensional generalized Laplacian and $\omega>0$ is a constant.
It is shown that $H_{\rm osc}$ is  self-adjoint,  
$\s(H_{\rm osc})=\sigma_{\rm disc}(H_{\rm osc})=\{\omega(n+\half )\}_{n=0}^\infty$ and each eigenvalue $\omega(n+\half)$ is simple. 
Since $\sum_{n=1}^{\infty}\frac{1}{(n+\half)^2}<\infty$, the assumption in  Proposition \ref{ag} holds. 
Hence $H_{\rm osc}$ has a time operator $T_{\rm osc}$  given by
$$
T_{\rm osc}f:=\frac{i}{\omega}\sum_{n=1}^{\infty}\left(\sum_{m\not=n}\frac{(e_m,f)}{n-m}\right)e_m,\quad f\in \D(T_{\rm osc}).
$$ 
One can show that $T$  is bounded and $\sigma(\overline{T})=[-\pi/\omega,\pi/\omega]$
(see \cite[Example 4.2]{am08a}).
}
\end{example}

\bc{tds}\label{cor-H-1}
Suppose  that 
 $\s (H)\setminus\{0\}=\s_{\rm disc} (H)=\{E_n\}_{n=1}^\infty$,  
each $E_n$ is simple,   
$E_1<E_2<\cdots <0$,  $0\not\in \s_{\rm p} (H)$,   and 
$\sum_{n=1}^\infty E_n^2<
\infty$. 
Then the operator $T_{\rm d}$ defined by
\begin{align}
\label{td11}
\td \phi:=i\sum_{n=1}^\infty 
 \lk \sum_{m\ne n}\frac{ (e_{m},  \phi)}{\frac{1}{E_n}-\frac{1}{E_m}}\rk 
e_{n},\quad \phi\in \D(\td):=\ms F
\end{align} 
is  a time operator of $H\f $,  
where $\ms F$ is given by \kak{F},  
i.e.,  
$[H\f,  \td]=-i\one$ on $\ms E$.  
\ec

\proof 
We see that $\sigma(H^{-1})=\s_{\rm disc} (H\f)=\{1/E_n\}_{n=1}^\infty$ and $\sum_{n=1}^{\infty}\frac{1}{(1/E_n)^2}<\infty$.
Hence the corollary  follows from  Proposition~\ref{ag}. 
\qed

\subsection{Case (II)}

In Corollary \ref{tds},  condition 
$\sum_{n=1}^\infty E_n^2<\infty$
is imposed to construct a  time operator of $H^{-1}$,  which is needed to apply  Proposition \ref{ag} with $H$ replaced by
$H^{-1}$. 
In this section, we show that 
the condition $\sum_{n=1}^\infty E_n^2<\infty$ can be
removed. The idea is to decompose $\hhh$ into  the direct sum 
of appropriate mutually orthogonal closed subspaces \cite{sw14}.

\begin{lemma}\label{lem-an-p} Let $p>1$ and $\{a_n\}_{n=1}^{\infty}$ be a complex sequence such that \linebreak $\lim_{n\to\infty}a_n=0$
and $a_n\not=a_m$ for $n\not=m$, $n,m\in \mathbb{N}$.
Let $A:=\{a_n|n\in \mathbb{N}\}$ be the set corresponding to the sequence $\{a_n\}_{n=1}^{\infty}$. Then there exist an $N \in \mathbb{N}\cup \{\infty\}$ 
and subsequences  $\{a_{kn}\}_{n=1}^{\infty}$ of $\{a_n\}_{n=1}^{\infty}$  ($k=1,\ldots,N$) such that
the sets 
$A_k:=\{a_{kn}|n\in \mathbb{N}\}$, $k=1,\ldots,N$, have the following properties:
\begin{align*}
&A_k\cap A_l=\emptyset\mbox{ for } k\not=l,\, k,l=1,\ldots,N;\\
&A=\cup_{k=1}^NA_k;\\
&\sum_{n=1}^{\infty}|a_{kn}|^p<\infty,\quad k=1,\ldots,N.
\end{align*}
\end{lemma}

\proof For each $k\in\mathbb{N}$, let $J_k:=\{a_n| 1/(k+1)<|a_n|\leq 1/k\}\subset A$ and
$\{k|J_k\not=\emptyset\}=\{k_1,k_2, \ldots\}$ with $k_1<k_2<\ldots$, which is an infinite set by the condition
$\lim_{n\to\infty}a_n=0$. It is obvious that $A=\cup_{n=1}^{\infty}J_{k_n}$ and
$J_{k_n}\cap J_{k_m}=\emptyset$ for all $(n,m)$ with $n\not=m$.
Let $a_{1n}\in J_{k_n}$. Then $\sum_{n=1}^{\infty}|a_{1n}|^p \leq \sum_{n=1}^{\infty}1/k_n^p<\infty$.
Let $A_1:=\{a_{1n}|n\in\mathbb{N}\}$ and $A':=A\setminus A_1$.
Write $A'=\{b_n|n\in\mathbb{N}\}$
with $b_n\not=b_m$ ($n\not=m$).
Then we can apply the preceding procedure on $\{a_n\}_{n=1}^{\infty}$ to $\{b_n\}_{n=1}^{\infty}$ to conclude that
there exists a subsequence $\{a_{2n}\}_{n=1}^{\infty}$ of $\{b_n\}_{n=1}^{\infty}$
such that $\sum_{n=1}^{\infty}|a_{2n}|^p<\infty$. Hence we obtain a subset
$A_2:=\{a_{2n}|n\in\mathbb{N}\}$. Obviously $A_1\cap A_2=\emptyset$. Then we give a similar consideration to
$A'':=A'\setminus A_2=A\setminus (A_1\cup A_2)$. In this way, 
by induction, we can show that, for each $k\in\mathbb{N}$,
there exists a subset $A_k$ which is empty or $A_k=\{a_{kn}|n\in\mathbb{N}\}\subset A$
such that $\sum_{n=1}^{\infty}|a_{kn}|^p<\infty$, $A_k\cap A_j=\emptyset, k\not=j$ and
$A=\cup_{k=1}^{\infty}A_k$ (if, for some $N\in\mathbb{N}$, $A=\cup_{k=1}^NA_k$, then
$A_k=\emptyset, k\geq N+1$). \qed

If a self-adjoint operator $S$ on a Hilbert space $\MSK$ is reduced by a closed subspace $\MSD$ of $\MSK$,
then we denote by $S_{\MSD}$ the reduced part of $S$ to $\MSD$, unless otherwise stated.

\begin{lemma}\label{sw}
Let 
 $\s (H)=\s_{\rm disc} (H)=\{E_n\}_{n=1}^\infty$,  %$\s (H)\setminus\{0\}\subset \s_{\rm disc} (H)$,   
 $E_1<E_2<\cdots<0$,  \linebreak $\lim_{n\to\infty} E_n=0$ and $0\not\in \s_{\rm p} (H)$.   
 Then there exist mutually orthogonal closed subspaces $\hhh_j$  of $\hhh$ ($j=1,\ldots,N,\, N\leq \infty$)
such that $\hhh$ is decomposed as  
$\hhh=\oplus_{j=1}^N \hhh_j$ ($N\leq \infty$) 
and (1)--(3) below are satisfied. 
\begin{list}{}{}
\item[(1)] Each $\hhh_j$ reduces $H$ and 
$\s (H_j)\setminus\{0\}=\s_{\rm disc}(H_j))=\{F_{jk}\}_{k=1}^\infty$, where $H_j:=H_{\hhh_j}$. 
\item[(2)] 
Each  eigenvalue $F_{jk}$ ($1\leq j\leq N,  1\leq k\leq \infty$)  is simple. 
\item[(3)] $\d \sum_{k=1}^\infty F_{jk}^2<\infty$ for each $1\leq j\leq N$. 
\end{list}
\end{lemma}
\proof
Note that $0$ is the unique accumulation point of the set $\{E_n|n\in \mathbb{N}\}$.   
Let $M_n$ be  the multiplicity of $E_n$ (which is finite, since $E_n\in \sigma_{\rm disc}(H)$). 
Let $\{e_n^i|i=1, \cdots, M_n\}$ be a complete orthonormal system (CONS) of $\ker(H-E_n)$:   
$He_n^i=E_n e_n^i,\,i=1,\ldots,M_n$.  
We set 
$$
\sup_{n\geq 1} M_n=M\quad \mbox{ and } \quad \limsup_{n\to\infty}M_n=m. 
$$
We consider two cases:  
 (A) $m=\infty$ and  (B) $m<\infty$.  

{\bf Case (A)}. Suppose that $m=\infty$.  
In this case,  $M=\infty$ and, 
for each  $k\geq 1$ and each $n$, there exists an $N\geq n$ 
such that $M_N\geq k$. 
Using this fact, we see that, for each $k\geq 1$, the  subspace 
$$
\ms G_k={\rm L. H.  } \{e_j^k| \ M_j\geq k\}
$$
is infinite dimensional and $\ms G_k$ is orthogonal to $\ms G_l$ for all $k,l$ with $k\not=l$.
Since $\{e_n^i|n\geq 1, i=1,\ldots,M_n\}$ is a CONS of $\hhh$, we have the orthogonal decomposition
\eq{gk3}
\hhh=\oplus_{k=1}^\infty \overline{\ms G_k}. 
\en

Fix $k$ and consider $\ms G_k$. 
Let $\ms A:=\s_{\rm disc} (H_{\ms G_k})=\{a_j|j\in\mathbb{N}\}$ ($=\{E_n|M_n\geq k\}$).
Then each eigenvalue $a_j$ is simple and $a_j\not=a_k$ for $j\not=k$, $\lim_{j\to\infty}a_j=0$.
Hence, we can apply Lemma \ref{lem-an-p} with $p=2$ to conclude that
 there exist an $N_k\leq \infty$ and subsets $\ms A_l:=
\{a_j^l\in \ms A|j\in\mathbb{N}, \sum_{j=1}^\infty |a_j^l|^2<\infty\}$ of $\ms A$
such that  $\ms A=\cup_{l=1}^{N_k}\ms A_l$ (a disjoint union).
Hence we can decompose $\overline{\ms G_k}$ as 
\eq{gk1}
\overline{\ms G_k}=\oplus_{l=1}^{N_k} \overline{\ms G_k^l}, 
\en
 where  $\ms G_k^l:={\rm L.H}\{g_j\in \ms G_k
 |Hg_j=a_j^lg_j, j\in \mathbb{N}\}$ (hence 
$\s_{\rm disc} (H_{\ms G_k^l})=\ms A_l$).  
Thus
\eq{gk4}
\hhh=\oplus _{k=1}^\infty \oplus_{l=1}^{N_k}  \overline{\ms G_k^l}
\en and the lemma follows.

{\bf Case (B)}. Suppose that $m<\infty$.  Then we have $m\leq M<\infty$. 
Hence we need only to consider four cases $ (a)- (d)$ below.

{\bf (a) $M=m=1$}. In this case, $\hhh=\overline{\ms G_1}$ 
and 
$\overline{\ms G_1}$ can be decomposed as \kak{gk1}.  
Then the lemma follows.

{\bf (b) $M\geq 2$ and $M=m$}. 
In this case, for all $k=1,\ldots,M$, $\ms G_k$ is infinite dimensional. 
Hence, in the same way as in  the case  $m=\infty$ 
we can see that $\hhh=\oplus_{k=1}^M\overline{\ms G_k}$  and 
$\overline{\ms G_k}$ can be decomposed as \kak{gk1}.  Thus the lemma follows.

{\bf (c) $M\geq 2$ and $m=1$}. 
In this case, there exists a $j_0\in \mathbb{N}$ such that for all $j\geq j_0$,  
$M_j=1$.  
Let $\ms B_k={\rm L. H.  } \{e_j^k|j< j_0,  k\leq M_j\}$,  $k=1, \cdots, M$ and
$\ms C:={\rm L. H.  } \{e_j^1|j\geq j_0\}$. 
Then we can decompose $\ms C$ as
$\overline{\ms C}=\oplus_{k=1}^{M} \overline{\ms C_k}$,  
where 
$\ms C_k={\rm L.H.}\{e_{j_k}^1|j_k \geq j_0,  j_k=j_0+k-1+Mr, \, r\in \{0\}\cup\mathbb{N}\}$ ($k=1,\ldots,M$).  
Define 
$\ms D_k=\ms B_k\oplus \overline{\ms C_k}$,  $k=1, \cdots, M$.  Then 
we have $\hhh=\oplus_{k=1}^M \ms D_k$.  
In the same way as in the  case (A),  we can decompose $\ms D_k$ like \kak{gk1}. Thus the lemma follows.

{\bf (d) $M\geq 2$,  $M>m$ and $m\geq 2$}. In this case, $\{j|M_j=m\}$ is a countable infinite set.
Hence, for $j=1,\ldots,m$,  $\ms G_j$ is infinite dimensional.
We have the orthogonal decomposition
\eq{gk2}
\hhh=\lk 
\oplus_{j=1}^{m-1}\overline{\ms G_j}\rk 
\oplus \ms K,
\en
where $\ms K=\lk \oplus_{j=1}^{m-1}\overline{\ms G_j}\rk^{\perp}$.  
The closed subspace $\ms K$ reduces $H$.
Since $\ms G_m\subset \ms K$, it follows that 
$\s (H_{\ms K})\setminus \{0\}=\sigma_{\rm disc}(H_{\ms K})$ is an infinite set. Let  $\sigma_{\rm disc}(H_{\ms K})=\{b_j\}_{j=1}^\infty$
and  $\beta_j$ be the multiplicity of eigenvalue $b_j$.  
Then $\sup_j\beta_j=M-m+1$ and 
$\sup_j \beta _j\geq \limsup _j\beta _j=1$.  Hence 
by 
 (a) and (c),  
we can decompose $\ms K$ 
as $\ms K=\oplus_{j=1}^{M-m+1}\ms K_j$,  where 
$\ms K_j$ is an infinite dimensional closed subspace of $\ms K$.  
Hence 
\eq{gk333}
\hhh=
\lk \oplus_{j=1}^{m-1}\overline{\ms G_j}\rk 
\oplus 
\lk \oplus_{j=1}^{M-m+1}\ms K_j\rk. 
\en
In the same way as in the case (A),  we can decompose $\overline{\ms G_j}$ and $\ms K_j$ like \kak{gk1}. Thus the lemma follows.  
\qed

Combining Corollary \ref{cor-H-1} and Lemma \ref{sw},  we can prove the following lemma.
  
\bt{wada+sasaki}\TTT{time operator of  $H\f$}\label{thm-TH}
Suppose  that 
 $\s (H)\setminus\{0\}=\s_{\rm disc} (H)=\{E_j\}_{j=1}^\infty$,  
$E_1<E_2<\cdots <0$, $\lim_{j\to\infty}E_j=0$,  and $0\not\in \s_{\rm p} (H)$.   
Then there exists a time operator $T_{-1}$ of  
$H\f$ with a dense CCR-domain for $(H\f,T_{-1})$. 
\et
\proof
By Lemma \ref{sw},   $\hhh$ can be decomposed as 
$\hhh=\oplus_{j=1}^N \hhh_j$ with $N\leq \infty$. 
By Proposition \ref{ag}, a  time operator $S_j$ of  $H\f_j$ exists:
$$[H\f _j,  S_j]=-i\one $$ on $\ms E_{j}:={\rm L. H.  }\{e_n^j-e_m^j,  n, m\in \mathbb N\}$,  where $\{e_n^j\}_{n=1}^\infty $ denotes the eigenvectors of $H_j$ such that 
$H_je_n^j=F_{jn}e_n^j$ and  
$\D (S_j)={\rm L. H. } \{e_n^j|n\in\mathbb N\}$.  
Define $T_{-1}$ by 
$T_{-1}:=\oplus_{j=1}^N S_j$ with 
$\D (T_{-1}):=\oplus_{j=1}^N \D (S_j)$ (algebraic direct sum).  
Then $T_{-1}$ is a time operator of  $H\f$ 
with a CCR-domain given by $\oplus_{j=1}^N \ms E_j$ (algebraic direct sum), which is dense in $\hhh$.     
\qed

\subsection{Case (III)}\label{subsec-III}

We next consider an extension of Proposition \ref{ag} to the case
where no restriction is imposed on the growth order  of the discrete eigenvalues $\{E_n\}_{n=1}^{\infty}$
of $H$.

\begin{lemma}\label{tsw}
Suppose  that  
 $\s (H)=\s_{\rm disc} (H)=\{E_n\}_{n=1}^\infty$  
 with $0<E_1<E_2<\cdots <E_n<E_{n+1}<\cdots $ and
 $\lim_{n\to\infty} E_n=\infty$.
 Then there exist mutually orthogonal closed subspaces $\hhh_j$ of $\hhh$ ($j=1,\ldots,N, N\leq \infty$) such that  
$\hhh=\oplus_{j=1}^N \hhh_j$ and 
(1)--(3) below are satisfied: 
\begin{list}{}{}
\item[(1)] Each $\hhh_j$ reduces $H$ and 
$\s (H_{\hhh_j})=\s_{\rm disc}(H_{\hhh_j})=\{F_{jk}\}_{k=1}^\infty$. 
\item[(2)] 
Each  $F_{jk}$ ($1\leq j\leq N,  k\in \mathbb{N}$)  is simple. 
\item[(3)] $\d \sum_{k=1}^\infty \frac{1}{F_{jk}^2}<\infty$ for each $1\leq j\leq N$. 
\end{list}
\end{lemma}
\proof 
Let $K=H^{-1}$. Then $K$ is self-adjoint and
$\s (K)\setminus\{0\}=\s_{\rm disc} (K)=\{1/E_n\}_{n=1}^\infty$,  %$\s (H)\setminus\{0\}\subset \s_{\rm disc} (H)$,   
 $1/E_1>1/E_2>\cdots>0$,  $\lim_{j\to\infty} 1/E_j=0$ and $0\not\in \s_{\rm p} (K)$.   
Hence, by applying  Lemma \ref{sw} to the case where $H$ and $E_n$ there are replaced by $-K$  and 
$-1/E_n$ respectively, we see that $\hhh$ has an orthogonal decomposition  
$\hhh=\oplus_{j=1}^N \hhh_j$ ($N\leq \infty$) 
with closed subspaces $\hhh_j$ of $\hhh$ 
such that (1)--(3) above  are satisfied. 
\qed
 
\bt{27}\TTT{time operator of  $H$}\label{thm-d-top}
Let 
 $\s (H)=\s_{\rm disc} (H)=\{E_n\}_{n=1}^\infty$,  
$E_1<E_2<\cdots $ and $\lim_{n\to\infty}E_n=\infty$.
Then there exists a  time operator $T$ of  $H$ with a dense CCR-domain for $(H,T)$. 
\et
\proof
The method of  proof is similar to that of Theorem \ref{wada+sasaki}. 
By Lemma \ref{tsw},   $\hhh$ can be decomposed as 
$\hhh=\oplus_{j=1}^N \hhh_j$ with $N\leq \infty$. 
By Proposition \ref{ag} a  time operator $T_{j}$ of  
$H_{\hhh_j}$ exists:  
$[H _{\hhh_j},  T_{j}]=-i\one $ on $\ms E_{j}={\rm L. H.  }\{e_n^j-e_m^j,  n, m\in \mathbb N\}$,  where $\{e_n^j\}_{n=1}^\infty $ denotes the eigenvectors of $H_{\ms H_j}$ such that 
$He_n^j=F_{jn}e_n^j$,  and the domain of $T_j$ is given by 
$\D (T_j)={\rm L. H. } \{e_n^j|n\in\mathbb N\}$.  
Define $T$ by 
$T:=\oplus_{j=1}^N T_j$ with 
$\D (T)=\oplus_j^N \D (T_j)$ (algebraic direct sum).  
Then $T$ is a  time operator of  $H$ with a CCR-domain $\oplus_{j=1}^N\ms E_j$ (algebraic direct sum), which is dense in $\hhh$. 
\qed

\begin{example}[$d$-dimensional quantum harmonic oscillator]
{\rm Let $\omega_j>0$ ($j=1,\ldots,d$) be a constant and  
$$
H_{{\rm osc},\,j}:=-\frac 12 D_j^2+\frac 12 \omega_j^2x_j^2
$$
acting in $L^2(\BR)$ (see Example \ref{ex-1hos}). 
Then the Hamiltonian of a $d$-dimensional quantum harmonic oscillator is given by
$$
H_{\rm osc}^{(d)}:=\sum_{j=1}^dH_{{\rm osc},\,j}
$$
acting in  $L^2(\BR)$.
It follows that $H_{\rm osc}^{(d)}$ is self-adjoint and
$$
\sigma(H_{\rm osc}^{(d)})=\sigma_{\rm disc}(H_{\rm osc}^{(d)})
=\left\{\sum_{j=1}^d\omega_j\left(n_j+\frac 12\right)|n_j\in \{0\}\cup\mathbb{N},\,j=1,\ldots,d\right\}.
$$
Hence, by Theorem \ref{thm-d-top}, $H_{\rm osc}^{(d)}$ has a time operator with  a dense CCR-domain.
}
\end{example}

\begin{example}[non-commutative harmonic oscillator]
{\rm 
Let $A$ and $J$ be $2\times 2$ matrices defined by 
$$A=\begin{pmatrix} \alpha& 0\\ 0& \beta\end{pmatrix},\quad
\alpha,\beta\geq0,\quad J=\begin{pmatrix} 0& {-1}\\ 1& 0\end{pmatrix}.$$
Let $\alpha\beta>1$. 
The Hamiltonian $H(\alpha,\beta)$ of the non-commutative harmonic oscillator \cite{p10} is defined by
the self-adjoint operator
\eq{ncho}
H(\alpha,\beta)=A\otimes (-\half \Delta+\half x^2)+J\otimes (xD+\half)
\en
on the Hilbert space
$\CC^2\otimes L^2(\RR)$, where $D$ is the generalized differential operator in $x$.
It is shown in \cite{iw07} that 
$\s(H(\alpha,\beta))=\s_{\rm disc}(H(\alpha,\beta))$ $=\{\lambda_n\}_{n=1}^\infty$ and 
the multiplicity of each $\lambda_n$ is not greater than $2$ with $\lambda_n\to\infty$ ($n\to\infty$). 
Hence, by Theorem \ref{thm-d-top}, $H(\alpha,\beta)$ has a time operator with a dense CCR-domain. 
}\end{example}

\begin{example}[Rabi model]{\rm 
Let $\mathbb{Z}_+:=\{0\}\cup\mathbb{N}$ be the set of nonnegative integers
and 
$$
\ell^2(\mathbb{Z}_+):=\left\{\psi=\{\psi_n\}_{n=0}^{\infty}\big|
\psi_n\in\CC,n\geq 0,\,\sum_{n=0}^{\infty}|\psi_n|^2<\infty\right\}
$$
be the Hilbert space of absolutely square summable complex sequences indexed by $\mathbb{Z}_+$.
The Hilbert space $\ell^2(\mathbb{Z}_+)$ is in fact the boson Fock space $\ms F_{\rm b}(\CC)$ over $\CC$ 
(e.g. \cite[p.53, Example 2]{rs1} and \cite[\S X.7]{rs2}): $\ell^2(\mathbb{Z}_+)=\ms F_{\rm b}(\CC)$.
We denote by  $a$  the annihilation operator on $\ms F_{\rm b}(\CC)$:
\begin{align*}
&(a\psi)_n:=\sqrt{n+1}\psi_{n+1}, n\geq 0, \\
&
\psi\in \D(a):=
\left\{\psi\in \ell^2(\mathbb{Z}_+)|\sum_{n=0}^{\infty}n|\psi_n|^2<\infty\right\}.
\end{align*}
We have $(a^*\psi)_0=0,\, (a^*\psi)_n=\sqrt{n}\psi_{n-1},\, n\geq 1$ for all $\psi\in \D(a^*)$.
The commutation relation $[a,a^*]=\one$ holds on the dense subspace $\ell_0(\mathbb{Z}_+)
:=\{\psi\in \ell^2(\mathbb{Z}_+)|\exists n_0\in \mathbb{N} \,\,\mbox{\rm such that}\,\, \psi_n=0,\, \forall n\geq n_0\}$. 

Let $\sigma_x,\sigma_y,\sigma_z$ be the  Pauli matrices:
\begin{align*}
 \sigma_x= \begin{pmatrix} 0 & 1 \\ 1 & 0 \end{pmatrix},\quad
 \sigma_y = \begin{pmatrix} 0 & -i \\ i & 0 \end{pmatrix},\quad
 \sigma_z = \begin{pmatrix} 1 & 0 \\ 0 & -1 \end{pmatrix}
\end{align*}
and 
\eq{rabi}
H_{\rm Rabi}:={\mu \s_z\otimes \one +\omega\one\otimes  \add a+ 
g \s_x\otimes (a+\add)}
\en
on $\CC^2\otimes \ms F_{\rm b}(\CC)$, where $\mu>0$, $\omega>0$ and $g\in \RR$ are constants.
The model whose Hamiltonian is given by $H_{\rm Rabi}$ is called the Rabi model \cite{rab36,rab37,bra11}. 
The matrix 
$$
U:=\frac 1{\sqrt{2}}(\sigma_x+\sigma_z)
$$
is unitary and self-adjoint. By direct computations using the properties that $\sigma_j\sigma_k+\sigma_k\sigma_j=2\delta_{jk},
j,k=x,y,z$, we see that
$$
\widetilde H_{\rm Rabi}:=UH_{\rm Rabi}U^{-1}=\mu\sigma_x+H,
\quad H:=\left(
\begin{array}{cc}
H_+ &0\\
0 & H_-
\end{array}
\right),
$$
where $H_{\pm}:=\omega a^*a\pm g(a+a^*)$ and
we have used the natural identification $\CC^2\otimes \ms F_{\rm b}(\CC)=\ms F_{\rm b}(\CC)\oplus \ms F_{\rm b}(\CC)$.
It is well known that the operator $\pi_g:=(g/\omega)\overline{i(a-a^*)}$ is self-adjoint and
$$
e^{\pm i\pi_g}ae^{\mp i\pi_g}=a\mp  \frac g{\omega}.
$$
Hence
$$
e^{\pm i\pi_g}H_{\pm}e^{\mp i\pi_g}=\omega a^*a - \frac {g^2}{\omega},
$$
implying that $\sigma(H_{\pm})=\sigma_{\rm disc}(H_{\pm})=\sigma(\omega a^*a-\frac{g^2}{\omega})=
\{\nu_n| n\in \mathbb{Z}_+\}$ with $\nu_n:= \omega n-\frac{g^2}{\omega}$. Hence
$\sigma(H)=\sigma_{\rm disc}(H)=\{\nu_n|n\in\mathbb{Z}_+\}$ with the multiplicity of each eigenvalue $\nu_n$ being two.
Since $\mu\sigma_x$ is bounded, it follows from the min-max principle that
$\widetilde H_{\rm Rabi}$ (and hence $H_{\rm Rabi}$) has purely discrete spectrum with $
\sigma(H_{\rm Rabi})=\sigma_{\rm disc}(H_{\rm Rabi})=
\sigma_{\rm disc}(\widetilde H_{\rm Rabi})=\{\nu_n'|n\in\mathbb{Z}_+\}$ 
satisfying $ \nu_n-\mu\leq \nu_{2n}'\leq \nu_n+\mu,\, n\geq 0$, where $\nu_0'\leq \nu_1'\leq \cdots \leq \nu_n'\leq
\nu_{n+1}'\leq \cdots$ counting multiplicities (see also \cite{bra11,mps14} for studies on spectral properties
of $H_{\rm Rabi}$). 
Hence we can apply  Theorem \ref{thm-d-top} to conclude that
$H_{\rm Rabi}$ has a time operator with a dense CCR-domain. 
}
\end{example}

\section{Ultra-Weak Time  Operators}\label{sec5}

\subsection{Ultra-weak time operators of a self-adjoint operator}

In this subsection, we consider the case where a self-adjoint operator $H$ obeys the assumption
of Theorem \ref{thm-TH} and ask if $H$ has a time operator.
We first give a formal heuristic argument.
By Theorem \ref{thm-TH}, we know that $H^{-1}$ has a time operator $T_{-1} $ with a dense CCR-domain 
for $(H^{-1},T_{-1})$.
Since the unique accumulation point of $\s (H)$ is $0$, but not $\infty$,  
it is not straightforward to apply Proposition \ref{ag} 
to construct a time operator of  $H$.  
The  key idea we use  is 
to regard $H$ as $H= (H\f)\f$.  
Let $f (x)=x^{-1}$. Then $H=f (H\f)$.  
Since $f' (x)=-x^{-2}$,  a formal application of 
Proposition  \ref{prop-hkm}
suggests that  
$A=-\half (T_{-1} H^{-2}+H^{-2} T_{-1})$
may be a time operator of $H$.
But, we note that no eigenvectors of $H$ are in $\D(H^{-2}T_{-1})$.
Hence it seems to be difficult to show  that $\D(A)\not=\{0\}$ and  
$\D(HA)\cap \D(AH)\not=\{0\}$. 
Thus we are led to consider a form version of $A$.

We use the notation  in the proof of Theorem \ref{thm-TH}. 
It is obvious that, for all  $k\in \mathbb{Z}$, 
$\D(S_j)\subset \D(H_j^k)$. Hence we define a sesquilinear form $\mft_j:\D(S_j)\times \D(S_j)\to\CC$ by
\begin{equation}
\mft_j[ \phi,\psi]:
=-\frac 12 \left\{(S_j\phi, H_j^{-2}\psi)+(H_j^{-2}\phi,S_j\psi)\right\},\quad
\phi,\psi\in \D(S_j). 
\end{equation}

\begin{lemma}\label{lem-uwt} Let
$$
H_j^{-1}\ms E_j:=\{H_j^{-1}\psi |\psi\in \ms E_j\}={\rm L.H.}\left\{\frac 1{E_n}e_n^j-\frac 1{E_m}e_m^j)\big|n,m\in\mathbb{N}\right\}.
$$ 
Then, for all $\psi,\phi\in H_j^{-1}\ms E$, $H_j\phi$ and $H_j\psi$ are in
$\D(S_j)$ and
\begin{equation}
\mft_j[ H_j\phi,\psi]-\mft_j[ \phi,H_j\psi]=-i(\phi,\psi).\label{TjH}
\end{equation}
\end{lemma}

\proof  Since $H_j(H_j^{-1}\ms E_j)=\ms E_j\subset \D(S_j)$,
$H_j\phi\in\D(S_j)$ for all $\phi\in H_j^{-1}\ms E_j$.
\linebreak By direct computations, we have
\begin{align*}
\mft[ H_j\phi,\psi]-\mft[ \phi,H_j\psi]
=&-\frac 12
\bigg\{
(S_jH_j\phi, H_j^{-2}\psi)
-(S_j\phi,H_j^{-1}\psi)\\
&+(H_j^{-1}\phi, S_j\psi)-(H_j^{-2}\phi, S_jH_j\psi)\bigg\}.
\end{align*}
We can write $\phi=H_j^{-1}\eta$ and $\psi=H_j^{-1}\chi$ with
$\eta,\chi\in \ms E_j$.
Then we have
\begin{align*}
(S_jH_j\phi, H_j^{-2}\psi)
-(S_j\phi,H_j^{-1}\psi)
&=(H_j^{-1}S_j\eta,H_j^{-1}\psi)-(S_jH_j^{-1}\eta, H_j^{-1}\psi)\\
&=
(-i\eta,H_j^{-1}\psi)=i(\phi,\psi),
\end{align*}
where we have used that $S_j$ is a time operator of $H_j^{-1}$ with a CCR-domain $\ms E_j$.
Similarly we have
\begin{align*}
(H_j^{-1}\phi, S_j\psi)-(H_j^{-2}\phi, S_jH_j\psi)
&=(H_j^{-1}\phi, S_jH_j^{-1}\chi)-(H_j^{-1}\phi, H_j^{-1}S_j\chi)\\
&=i(\phi, \psi).
\end{align*}
Thus (\ref{TjH}) follows.\qed

Lemma \ref{lem-uwt} shows that $\mft_j$ is an ultra-weak time operator of $H_j$ with 
$H_j^{-1}\ms E_j$ being an ultra-weak CCR-domain.

We introduce
$$
\widetilde{\ms E}:=\oplus_{j=1}^NH_j^{-1}\ms E_j \,\,\,\, (\mbox{algebraic direct sum}).
$$
Since $H_j^{-1}\ms E_j$ is dense in $\hhh_j$, $\widetilde {\ms E}$ is dense in
$\hhh$ and $\widetilde {\ms E}\subset \D(H)$.

\begin{theorem}\label{thm-uwt} Under the same assumption as in Theorem \ref{thm-TH},
there exists an ultra-weak  time operator $\mft_{\rm p}$ of  
$H$  with $\widetilde {\ms E}$ being an ultra-weak CCR-domain. 
\end{theorem}

\proof Let $T_{-1}$ be as in Theorem \ref{thm-TH} and  define a sesquilinear form $\mft_{\rm p}:
\D(T_{-1})\times \D(T_{-1})\to\CC$ by
$$
\mft_{\rm p}[ \psi,\phi]:=
\sum_{j=1}^N\mft[ \psi_j,\phi_j], \quad \psi=(\psi_j)_{j=1}^N, \phi=(\phi_j)_{j=1}^N\in \D(T_{-1}).
$$
We remark that, in the case $N=\infty$, $\psi_j=0$ for all sufficiently large $j$ and hence
the sum $\sum_{j=1}^N$ on the right hand side is over only finite terms, being well defined.
It follows from  Lemma \ref{lem-uwt} that, for all $\psi,\phi\in \widetilde {\ms E}$,
$H\psi,H\phi\in{\ms E}\subset \D(T_{-1})$ and
$$
\mft_{\rm p}[ H\phi,\psi]-\mft_{\rm p}[ \phi, H\psi]=-i(\phi,\psi).
$$
This means that $\mft_{\rm p}$ is an ultra-weak time operator of $H$ with
$\widetilde {\ms E}$ being an ultra-weak CCR domain.   
\qed

We now proceed to showing
existence of an ultra-weak time operator
of a self-adjoint operator in a general class.

\begin{definition}[class $S_1(\hhh)$]\label{def-S1}
{\rm A self-adjoint operator
$H$ on $\hhh$ is said to be in the class $S_1(\hhh)$ if it has the following properties (H.1)--(H.4):
\begin{list}{}{}
\item[(H.1)]
$\s_{\rm sc} (H )=\emptyset$.
\item[(H.2)]
$\s_{\rm ac} (H)=[0, \infty)$.
\item[(H.3)]
$\s_{\rm disc} (H)=\sigma_{\rm p}(H)=\{E_n\}_{n=1}^\infty$, 
$E_1<E_2<\cdots<0$,  $\lim_{n\to\infty}E_n=0$ (hence $0\not\in \s_{\rm p} (H)$). 

\item[(H.4)]
There exists a strong time operator $T_{\rm ac}$ 
of  $H_{\rm ac}$ in  
$\ms H_{\rm ac}(H)$.  
\end{list}
}
\end{definition}

Let $H\in S_1(\hhh)$. Then we have the orthogonal decomposition
$$
\hhh=\hhh_{\rm ac}(H)\oplus \hhh_{\rm p}(H).
$$ 
By (H.3), we can apply  Theorem \ref{thm-uwt}
to the case where $H$ is replaced by $H_{\rm p}$ to conclude that
$H_{\rm p}$ has an ultra-weak time operator $\mft_{\rm p}$ with a dense ultra-weak CCR domain $\ms E_{\rm p}$
such that
$$
\mft_{\rm p}[ H_{\rm p}\phi,\psi]-\mft_{\rm p}[ \phi,H_{\rm p}\psi]=-i(\phi,\psi),\quad \phi,\psi\in \ms E_{\rm p}.
$$
We denote by $\MSD_{\rm p}$ the subspace $\D(T_{-1})$ in the proof
of Theorem \ref{thm-uwt}. Hence $\mft_{\rm p}:\MSD_{\rm p}\times \MSD_{\rm p}\to\CC$ with
$\ms E_{\rm p}\subset \D(H_{\rm p})\cap \MSD_{\rm p}$ and $H_{\rm p}\ms E_{\rm p}\subset \MSD_{\rm p}$.
By (H.4), there exists a dense CCR-domain  $\MSD_{\rm ac}$ for $(H_{\rm ac}, T_{\rm ac})$.
Let $\widetilde {\ms E}_{\rm p}:=H_{\rm p}^{-1}{\ms E}_{\rm p}$ and
\begin{equation}
\MSD_H:=\MSD_{\rm ac}\oplus \widetilde{\ms E}_{\rm p},
\end{equation}
which is dense in $\hhh$.

We define a sesquilinear form 
$\mft_H  :(\hhh_{\rm ac}(H)\oplus  \MSD_{\rm p})\times (\D(T_{\rm ac})\oplus \MSD_{\rm p})\to\CC$ 
by   
\begin{align}
\mft_H [\phi_1\oplus\phi_2,  \psi_1\oplus\psi_2]  =& 
(\phi_1,  T_{\rm ac}\psi_1)+\mft_{\rm p}[  \phi_2, \psi_2],\nonumber\\
& \quad \phi_1\in \hhh_{\rm ac}(H),
\psi_1\in \D(T_{\rm ac}), \phi_2,\psi_2\in \MSD_{\rm p}.\label{time1} 
\end{align}
Now we are in the position to state and prove the main result
in this section.

\begin{theorem}[abstract ultra-weak time operator 1]\label{thm-uwt1}Let $H\in S_1(\hhh)$.
Then the sesquilinear form $\mft_H$ defined by (\ref{time1})
is an ultra-weak time operator of $H$ with $\MSD_H$ being an ultra-weak CCR-domain.
\end{theorem}

\proof Let $\mft_{\rm ac}:\hhh_{\rm ac}(H)\times \D(T_{\rm ac})\to\CC$
by 
$$\mft_{\rm ac}[\phi,\psi]:=(\phi,T_{\rm ac}\psi),\,\phi\in \hhh_{\rm ac}(H),\psi\in\D(T_{\rm ac}).$$
Then, by Remark \ref{rem-uwt}-(2), $\mft_{\rm ac}$ is an ultra-weak time operator
of $H_{\rm ac}$ with $\MSD_{\rm ac}$ being an ultra-weak CCR-domain.
Then, in the same way as in the proof of Theorem \ref{thm-uwt},
one can show that $\mft_H$ is an ultra-weak time operator of $H$ with
$\MSD_H$ being an ultra-weak CCR-domain.
  \qed

\subsection{Ultra-weak time operators of $f(H)$}

We can also construct an ultra-weak time operator  of  
$f(H)$ for some function $f:\RR\to\RR$. 
A strong time operator of $f(H_{\rm ac})$ is already constructed in Proposition \ref{hkm09}.
Hence we need only to construct an ultra-weak time operator 
of  $f(H_{\rm p})$.  
A set of conditions for that  is as follows.  
\begin{assumption}\label{hara}
{\rm Let $H\in S_1(\hhh)$.
\begin{list}{}{}  
 \item{(1)}  The function $f:\RR\to\RR$ satisfies the same 
assumption as in Proposition~\ref{hkm09}. 
 \item{(2)} The function $f$ is continuous at $x=0$. 
 \item{(3)} $f(\s_{\rm disc} (H))$ is an infinite  set  such that
 the multiplicity of each point in $f(\s_{\rm disc}(H))$ as an eigenvalue of $f(H)$ is finite.
\item{(4)} $f (0)\not\in \s_{\rm p} (f (H))$. 
 \end{list}
}\end{assumption}
Suppose that Assumption \ref{hara} holds.
Then  $f(H)$ is self-adjoint and reduced by $\hhh_{\rm ac}(H)$ and $\hhh_{\rm p}(H)$
(these properties  follow from only the fact that  $f:\RR\to\RR$, Borel measurable, and the general theory of functional calculus).
We denote the reduced part of $f(H)$ to $\hhh_{\rm ac}(H)$ and $\hhh_{\rm p}(H)$
by $f(H)_{\rm ac}$ and $f(H)_{\rm p}$ respectively.
By the functional calculus,
we have 
$f (H)_{\rm ac}=f (H_{\rm ac})$, $f (H)_{\rm p}=f (H_{\rm p})$ and
$f (H)=f (H_{\rm ac})\oplus f (H_{\rm p})$.
This implies that 
\begin{equation}
\sigma(f(H))=\sigma(f(H_{\rm ac}))\cup\sigma(f(H_{\rm p}))=\ov{f ([0, \infty))}\cup \ov{\{f (E_j)\}_{j=1}^\infty}.
\end{equation}

\bc{main3}
\TTT{abstract ultra-weak  time  operator 2}
Under Assumption \ref{hara}, there exists an ultra-weak time operator $\mft_{H}^f$  of  
$f (H) $ with a dense ultra-weak CCR-domain.
\ec

\proof Let $T_{\rm ac}$ be a strong time operator of  $H_{\rm ac}$.  
Then the strong time operator of
 $f (H_{\rm ac})$ is given by 
$$T_{{\rm ac}}^f=\half\ov{ (T_{\rm ac}f' (H_{\rm ac})^{-1}
+f' (H_{\rm ac})^{-1}T_{\rm ac}) \lceil D}$$ by 
Proposition \ref{hkm09}, where 
$D:=\{g(H_{\rm ac}) \D (T_{\rm ac})| g\in C_0^\infty (\RR\setminus L\cup K)\}$.  
Let $$\tilde f (x)=f (x)-f (0). $$
Then $\lim_{x\to 0}\tilde f (x)=0$.  We can write
$\s (\tilde f (H_{\rm p}))=\{F_j\}_{j=1}^\infty$, where $F_j\not=F_k,\,j\not=k$ and  
the multiplicity of each $F_j$ is finite. It follows that $\lim_{j\to\infty}F_j=0$  
and 
$0\not\in \s_{\rm p} (\tilde f (H_{\rm p}))$.  
Hence, by a minor modification of the proof of Lemma \ref{lem-uwt},   
we can show that 
there is an ultra-weak time operator $\mft_{\rm p}^{\tilde f}:\MSD_{\rm p}^f\times \MSD_{\rm p}^f\to \CC$ of  
$\tilde f (H_{\rm p})$, where $\MSD_{\rm p}^f$ is a dense  subspace in $\hhh_{\rm p}(H)$.  
We define  a sesquilinear  form 
$\mft_{H}^f :(\hhh_{\rm ac}(H)\oplus \MSD_{\rm p}^f)\times (\D(T_{\rm ac}^f)\oplus
\MSD_{\rm p}^f)\to\CC$ 
 by   
\eq{time11}
\mft_H^f [\phi_1\oplus\phi_2,  \psi_1\oplus\psi_2]    = (\phi_1,  T_{\rm ac}^f\psi_1)+\mft_{\rm p}^{\tilde f}
 [  \phi_2, \psi_2]   
\en
for 
$\phi_1\in \hhh_{\rm ac}(H), \psi_1\in \D (T_{\rm ac}^f)$ and $\phi_2, \psi_2\in \MSD_{\rm p}^f$.  
Note that $f (0)$ is a scalar. Then one can show that 
$\mft_H ^f$ is an ultra-weak time operator with 
$\MSD_{\rm ac}\oplus\widetilde {\ms E}_{\rm p}^{\tilde f}$ being
a  ultra-weak CCR-domain, where $\MSD_{\rm ac}$ is a dense  CCR-domain for $(f(H_{\rm ac}), T_{\rm ac}^f)$.
and $\widetilde{\ms E}_{\rm p}^{\tilde f}$ is an ultra-weak CCR-domain for $(\tilde f(H_{\rm p}),\mft_{\rm p}^{\tilde f})$.
\qed

\section{Applications to Schr\"odinger Operators}\label{sec6}

\subsection{Ultra-weak time operators of Schr\"odinger operators}\label{ex}

In this subsection, we apply Theorem \ref{thm-uwt1}
to the Schr\"odinger operator $\HS$ given by (\ref{HV}) to show that,
for a general class of potentials $V$, $\HS$ has an ultra-weak time operator with a {\it dense} ultra-weak CCR-domain.
This is done by  collecting known results on spectral properties of Schr\"odinger operators.   

Suppose that $V$ is of the form 
\eq{agmon}
V (x)=\frac{W (x)}{ (|x|^2+1)^{\han +\eps}},
\en
where $\eps>0$ and    
$W:\BR\to\RR$ is  a Borel measurable function  such that 
 $W (-\Delta+i)\f$ is a compact operator on $\LR$. 
Such a potential  $V$  is called an {\it Agmon potential} (\cite[p.439]{rs3} or \cite[p.169]{rs4}).  
It is easily shown that  $V$ is relatively compact with respect to 
the free Hamiltonian $H_0$ given by (\ref{H0}).
Hence, by a general fact \cite[p.113, Corollary 2]{rs4},  
$\HS$ is self-adjoint with $\D(\HS)=\D (H_0)$ and
\begin{equation}
\s_{\rm ess} (\HS )=\s_{\rm ess} (H_0)=[0, \infty),\label{HV-ess}
\end{equation}
where, for a self-adjoint operator $S$, $\s_{\rm ess}(S)$ denotes the essential spectrum of $S$.

Following facts are known as 
Agmon-Kato-Kuroda theorem:
\bp{akk}\TTT{absence of $\s_{\rm sc} (H)$, existence and completeness of wave operators}
Let $V$ be an Agmon potential. 
Then:
\begin{list}{}{}  
\item{(1)} $\s_{\rm sc} (\HS )=\emptyset$. 
\item{(2)}
The set of positive eigenvalues of $\HS $ is a discrete subset of $ (0, \infty)$. 
\item{(3)} The wave operators 
$\d \Omega_{\pm} :={\rm s}\!-\!\lim_{t\to\pm\infty}
e^{it\HS }e^{-itH_0}$ 
exist and complete: ${\rm Ran}(\d\Omega_{\pm})=\hhh_{\rm ac}(\HS)$. 
In particular $\s_{\rm ac} (\HS )=[0,\infty)$.  
\end{list}
\ep
\proof
See  \cite[Theorem XIII. 33]{rs4}. 
\qed

In order to construct an ultra-weak time operator of  
$(\HS)_{\rm p}$, we need 
the condition 
$\#\s_{\rm disc}(\HS) =\infty$. 
For this purpose, we introduce an assumption. 
\begin {assumption}\label{v2}
{\rm There are constants $R_0, a>0$ and $\delta>0$ such that
\eq{takusan}
V (x)\leq -\frac{a}{|x|^{2-\delta}}\quad \mbox{ for } |x|>R_0.
\en
}
\end{assumption}

\begin{lemma}\TTT{infinite number of negative eigenvalues}
Let $V$ be an Agmon potential. Then, under Assumption \ref{v2},
$\s_{\rm disc} (\HS)\subset (-\infty, 0)$ and 
$\s_{\rm disc} (\HS)$ is an infinite set. 
In particular, the point $0\in\RR$ is the unique accumulation point of  $\s_{\rm disc}(\HS)$.
\end{lemma}

\proof Let $\mu_1:=\inf_{\psi\in \D(\HS);\|\psi\|=1}(\psi,\HS\psi)$ and
$$
\mu_n:=\sup_{\phi_1,\ldots,\phi_{n-1}\in \LR}\inf_{\substack{\psi\in D(\HS);\|\psi\|=1\\
\psi\in \{\phi_1,\ldots,\phi_{n-1}\}^{\perp} }}(\psi,\HS\psi),\quad n\geq 2.
$$
In the case $d=3$, it is already known that $\mu_n<0$ for all $n\in \mathbb{N}$ \cite[Theorem XIII.6(a)]{rs4}.
It is easy to see that the method of the proof of this fact is valid also in the case of arbitrary $d$.
Hence we have $\mu_n<0$ for all $n\in \mathbb{N}$.
Then (\ref{HV-ess}) and the min-max principle imply the desired results.
\qed

\begin{assumption}\label{v3}
{\rm The potential  $V$  is spherically symmetric, $V=V(|x|)$,  and 
\eq{nai}
\int_a^\infty |V (r)| dr<\infty
\en
for some $a>0$. 
}\end{assumption}

\begin{lemma}\TTT{absence of strictly positive 
eigenvalues} Let $V$ be an Agmon potential. Then,
under Assumption \ref{v3}, 
$\HS$ 
has no strictly positive eigenvalues.  
\end{lemma}

\proof Since $\D(V)\supset \D(H_0)\supset C_0^{\infty}(\BR)$,
it follows that $V\in L_{\rm loc}^2(\BR\setminus \{0\})$.
Hence we can apply  \cite[Theorem XIII.56]{rs4} to 
derive the desired result.\qed

\bt{main2}\label{thm-uwt2}
Let $V$ be an Agmon potential such that $0\not\in \s_{\rm p} (\HS) $. 
Suppose that Assumptions \ref{v2} and \ref{v3} hold. 
Then   $\HS$ has an ultra-weak time operator with a dense ultra-weak CCR-domain.  
\et
\proof
By Proposition \ref{akk},  $\s_{\rm sc} (\HS)=\emptyset$ and 
the wave operators $\Omega_{\pm}$ exist and are complete. 
Hence, by Theorem \ref{thm-st-uni1}, 
$$
T_{{\rm ac}, j\pm}:=\Omega_{\pm} \tau_j\Omega_{\pm}^{-1}
P_{\rm ac}(\HS)\quad (j=1,\ldots,d)
$$ 
are strong time operators of  $(\HS)_{\rm ac}$, where
$\tau_j:=\widetilde T_{{\rm AB},j}$ or $T_{{\rm AB},j}'$ denotes the Aharonov-Bohm time operators  in Example \ref{ex-AB}.
Under  Assumptions \ref{v2} and \ref{v3}, 
we can see that 
$\s (\HS )=\{E_j\}_{j=1}^\infty\cup[0, \infty)$,  
$E_1<E_2<\cdots<0$, $\lim_{n\to\infty}E_n=0$,   
$\s_{\rm disc}(\HS )=\{E_n\}_{n=1}^{\infty}$ and
$\s_{\rm ac} (\HS )=[0, \infty)$.  Hence $\HS\in S_1(L^2(\BR))$.
Thus, by Theorem \ref{thm-uwt1}, we obtain the desired result.  
\qed

Finally we consider conditions for   
the absence of zero eigenvalue of 
$\HS$.

\bp{v4}\TTT{absence of zero eigenvalue}
Assume the following (1) and (2):
\begin{list}{}{}
\item{(1)} $d\geq 3$, $V\in  L_{\rm loc}^{d/2}(\BR)$. 
\item{(2)} $V$ can be written as $V=V_1+V_2$, where $V_1$ and $V_2$ are real-valued 
Borel measurable functions on $\BR$ satisfying the following conditions: 
\begin{list}{}{}
\item[(i)] There exists a constant $R>0$ such that $V_1$ and $V_2$ are locally bounded on  $S_R=\{x\in\BR | |x|>R\}$
 and $V_1$ is strictly negative on 
$S_R$,  
\item[(ii)] 
Let $S^{d-1}:=
\{w\in\BR|\,|w|=1\}$, the $(d-1)$-dimensional unit sphere.  
Then $V_1(rw)$ ($r=|x|$) is differentiable in $r>R$ and
there exist a constant $s\in (0,1)$ and a positive  differentiable  function $h$
on $[R,\infty)$ such that
$$
\d \sup_{w\in S^{d-1}} \frac{d}{dr}(r^{s+1} V_1(rw))\leq -r^{s} h(r)^2,\quad r>R.
$$ 
\item[(iii)] $\displaystyle\lim_{r\to\infty}\d \frac{r^{-1}+r \sup_{w\in S^{d-1}} |V_2(rw)|}{h(r)}=0$.
\item[(iv)] There exists a constant $C>0$ such that $\d \frac{d}{dr} h(r)\leq Ch^2(r)$ on $S_R$.
\item[(v)] For all all $f\in D(\HS)$, 
$$\int_{S_R} h^2(|x|)|f(x)|^2 dx<\infty,\quad 
\int_{S_R} |V_1(x)| |f(x)|^2 dx<\infty.
$$
\end{list}
\end{list}
Then $0\not\in \s_{\rm p}(\HS)$.
\ep
\proof
This is due to \cite[Theorem 2.4]{fs04} and \cite{jk85}.  Also see 
\cite{uch87}. 
\qed

A key fact to prove  Proposition \ref{v4} is as follows.  Condition 
$d\geq3$ and  
$V\in L_{\rm loc}^{d/2} (\BR)$ imply 
that, if a solution $f$ of partial differential equation 
 $-\Delta f+Vf=0$ satisfies that $f(x)=0$ for all $x\in S_R$ with some $R>0$,  
 then $f (x)=0$ for all  $x\in \BR$ by the unique continuation proven 
 in \cite{jk85}.  

\begin{example}\label{ex-hyd}{\rm Let $d\geq 3$ and $V (x)=-1/|x|^{2-\eps}$ with  
$0<\eps<2$.  Then it is easy to check that the potential  $V$ satisfies
 conditions (1) and (2) in Proposition \ref{v4} (take $V_1=V$, $V_2=0$ and $h(r)=
\sqrt{s-1+\eps}\,r^{(\eps-2)/2}, \, r>0$ with $1-\eps<s<1$).  Hence, by Proposition \ref{v4}, $\HS $ has 
no zero eigenvalue.
In particular, 
the hydrogen Schr\"odinger operator 
\begin{equation}
H_{\rm hyd}:= H_0-\frac{\gamma}{|x|} \label{H-hyd}
\end{equation}
for $d=3$ with a constant $\gamma>0$ 
has no zero eigenvalue.}  
\end{example}

\begin{example}\label{47}
{\rm Let $d\geq 3$.  
Suppose that $U\in L^\infty (\RR^3)$.  
Then 
$$V (x)=\frac{U (x)}{ (1+|x|^2)^{\han +\eps}}$$ is an Agmon potential for all $\eps>0$.  
Suppose that 
$U$  is negative,  continuous,  spherically symmetric and 
satisfies that 
$U (x)=-1/|x|^\alpha$ for $|x|>R$  
with $0<\alpha<1$ and $R>0$. 
For each $\alpha$,  we 
can choose $\eps>0$ such that 
$2\eps+\alpha<1$. 
Hence $V$ satisfies \kak{takusan} and \kak{nai}.
Moreover it is easy to see that  $V$ satisfies (1) and (2) in Proposition \ref{v4}
with 
$$
V_1(x):=-\frac {\chi_{[R,\infty)}(|x|)}{|x|^{1+2\eps+\alpha}},\quad
V_2(x):=\frac{U(x)}{(1+|x|^2)^{\han+\eps}}+\frac {\chi_{[R,\infty)}(|x|)}{|x|^{1+2\eps+\alpha}},
$$
where $\chi_{[R,\infty)}$ is the characteristic function of the interval $[R,\infty)$.
Hence, by Proposition \ref{v4},  $0\not\in \s_{\rm p}(\HS )$.
Thus, by Theorem \ref{thm-uwt2}, 
$\HS$ has an ultra-weak time operator with
a dense ultra-weak CCR-domain.  
}
\end{example}

\begin{example}
\TTT{hydrogen atom}\label{ex-hyd2}
\label{main4}
{\rm 
It is known that the hydrogen Schr\"odinger operator  
$H_{\rm hyd}$ given by (\ref{H-hyd}) is self-adjoint with
$\D(H_{\rm hyd})=D(H_0)$. 
It is easy to see that the  Coulomb potential $-\gamma/|x|$ with $d=3$ is not an Agmon potential.
Hence we can not apply Theorem \ref{thm-uwt2} to the case $\HS=H_{\rm hyd}$.
But we can show that $H_{\rm hyd}$ has an ultra-weak time operator in the following way.
The spectral properties of $H_{\rm hyd}$ are also well known:
$$
\s(H_{\rm hyd})=\sigma_{\rm p}(H_{\rm hyd})\cup \sigma_{\rm ac}(H_{\rm hyd}), \quad \sigma_{\rm sc}(H_{\rm hyd})=\emptyset
$$
with
$$
\sigma_{\rm p}(H_{\rm hyd})=\sigma_{\rm disc}(H_{\rm hyd})=
\left\{-\frac {m\gamma^2}{2n^2}|n\in\mathbb{N}\right\},\quad \sigma_{\rm ac}(H_{\rm hyd})=[0,\infty).
$$
The fact  that
$0\not\in \s_{\rm p}(H_{\rm hyd})$ follows from Example \ref{ex-hyd} and Proposition \ref{v4}.   
It is shown that the modified wave operators 
$\mbox{\rm s-}\lim_{t\rightarrow \pm\infty} e^{itH_{\rm hyd}}Je^{-itH_0}$ 
with  some unitary operator $J$ exist and are complete  \cite[Theorems XI. 71 and XI.72]{rs3}. 
These facts imply that $H_{\rm hyd} \in S_1(L^2(\RR^3))$.
Thus, by Theorem \ref{thm-uwt1}, $H_{\rm hyd}$ has an ultra-weak time operator with a dense ultra-weak CCR-domain. 
}\end{example}

\subsection{Ultra-weak time operators of $f(\HS)$}

In this subsection, we assume that $\HS\in S_1(L^2(\BR))$ (see Definition \ref{def-S1})
and give some examples of functions $f:\RR\to\RR$ such that $f(\HS)$  has a ultra-weak-time operator
with a dense ultra-weak CCR-domain.
We first give a sufficient condition for (4) in Assumption \ref{hara} to hold. 

\begin{lemma}\label{lem-f0} Let $\HS\in S_1(L^2(\BR))$ and  $f:\RR\to\RR$, Borel measurable.
Suppose that,  for all $n\in \mathbb{N}$, $f(E_n)\not=f(0)$ and $f(x)\not=f(0),\,{\rm a.e.}x\geq 0$.
Then  $f(0)\not\in\sigma_{\rm p}(\HS)$.
\end{lemma}

\proof Let $\psi\in D(f(\HS))$ such that $f(\HS)\psi=f(0)\psi$. Then 
$$\|(f(\HS)-f(0))\psi\|^2=0$$ which is equivalent to
$\int_{\RR}|f(\lambda)-f(0)|^2d\|E(\lambda)\psi\|^2=0$, where $E(\cdot)$ is the spectral measure of $\HS$.
We can decompose $\psi$ as $\psi=(\psi_{\rm ac},\psi_{\rm p})\in \hhh_{\rm ac}(\HS)\oplus\hhh_{\rm p}(\HS)$.
We denote by $\rho$ the Radon-Nykod\'ym derivative of the absolutely continuous
 measure $\|E(\cdot)\psi_{\rm ac}\|^2$ with respect to the Lebesgue measure on $\RR$.
Then we have
\begin{align*}
\int_{\RR}|f(\lambda)-f(0)|^2d\|E(\lambda)\psi\|^2&=
\sum_{n=1}^{\infty}|f(E_n)-f(0)|^2\|E(\{E_n\})\psi_{\rm p}\|^2\\
& \quad +
\int_{[0,\infty)}|f(\lambda)-f(0)|^2\rho(\lambda)d\lambda.
\end{align*}
Hence, by the present assumption, $\|E(\{E_n\})\psi_{\rm p}\|^2=0\cdots (*)$ for all $n\in \mathbb{N}$ and
$\int_{[0,\infty)}|f(\lambda)-f(0)|^2\rho(\lambda)d\lambda=0\cdots(**)$.
Equation $(*)$ implies that $E(\{E_n\})\psi_{\rm p}=0,\,\forall n\geq 1$.
Since $\HS$ is  $S_1(L^2(\BR))$, it follows  that $\psi_{\rm p}\in \hhh_{\rm p}(\HS)^{\perp}$. 
Hence $\psi_{\rm p}=0$. On the other hand, $(**)$ implies that $\rho(\lambda)=0$ a.e.$\lambda\in [0,\infty)$, from
which it follows that $\psi_{\rm ac}=0$. Thus $\psi=0$.
\qed

\begin{theorem}\label{thm-uwt-HS} Let $\HS\in S_1(L^2(\BR))$ and $f:\RR\to\RR$, Borel measurable.
Assume  the following (1)--(4): 
\begin{list}{}{}
\item{(1)}  The function $f:\RR\to\RR$ satisfies the same 
assumption as in Proposition~\ref{hkm09}. 
 \item{(2)} The function $f$ is continuous at $x=0$. 
 \item{(3)} $f(\s_{\rm disc} (\HS))$ is an infinite set  such that
 the multiplicity of each point in $f(\s_{\rm disc}(\HS))$ as an eigenvalue of $f(\HS)$ is finite.
\item{(4)}  For all $n\in \mathbb{N}$, $f(E_n)\not=f(0)$ and $f(x)\not=f(0),\,{\rm a.e.}x\geq 0$.
\end{list}
Then $f(\HS)$ has an ultra-weak time  operator  with a dense ultra-weak CCR-domain.
\end{theorem}

\proof By Lemma \ref{lem-f0}, property (4) in Assumption \ref{hara}
is satisfied. Hence, by Corollary \ref{main3}, the desired result is derived.
\qed

In Examples below, we assume that $\HS\in S_1(L^2(\BR))$.

\begin{example}\TTT{$f (\HS)=e^{-\beta \HS}$}\label{exp}
{\rm 
Let $f (x)=e^{-\beta x}$,  $\beta\in\RR\setminus\{0\}$. 
Then it is easy to see that the function $f$ satisfies the assumption in Theorem \ref{thm-uwt-HS}. 
Hence $e^{-\beta \HS}$ has an ultra-weak time  operator  with a dense ultra-weak CCR-domain. 
Note that, if $\beta>0$ (resp. $\beta <0$), $e^{-\beta \HS}$ is bounded (resp. unbounded). 
In particular,  
$e^{-\beta H_{\rm hyd}}$ has   an ultra-weak time  operator  with a dense ultra-weak CCR-domain. 
}\end{example}

\begin{example}\TTT{$f (\HS)=\sum_{j=0}^N a_j H_V^j$}\label{poly}
{\rm 
Let $f (x)=\sum_{j=0}^N a_j x^j$ be a real polynomial ($a_j\in \RR, N\in \mathbb{N}, \,a_N\not=0$). 
We have $f(0)=a_0$. 
Suppose that, for all $n\in \mathbb{N}$, $\sum_{j=1}^Na_jE_n^j\not=0$ and $\sum_{j=1}^Na_jx^{j-1}\not=0,\,x\geq 0$.
Then one can show that $f$ satisfies the assumption in Theorem \ref{thm-uwt-HS}. 
Hence $\sum_{j=0}^Na_j\HS^j$ has an ultra-weak time  operator  with a dense ultra-weak CCR-domain. 
In particular,  
$\sum_{j=0}^Na_jH_{\rm hyd}^j$  has  an ultra-weak time  operator  with a dense ultra-weak CCR-domain. 
}\end{example}

\begin{example}\TTT{$f (\HS)=\sin (2\pi \beta \HS)$}\label{411}
{\rm 
Let $f (x)=\sin (2\pi \beta  x) $,  $\beta\in \RR\setminus\{0\}$. 
Then $f(0)=0$. Let  $\beta\not\in \{k/2E_n|k\in \mathbb{Z}, n\in\mathbb{N}\}$. Then
$\sin (2\pi\beta E_n)\not=0$ for all $n\in \mathbb{N}$ and hence
$f(E_n)\not=f(0)$. It is obvious that $f(x)\not=f(0)$ for a.e.$x\geq 0$.
Moreover $\Lambda:=\{\sin(2\pi\beta E_n)|n\in \mathbb{N}\}$ is an infinite  set 
and  each point  in $\Lambda$ as an eigenvalue of $\sin(2\pi\beta \HS)$  is
in $\sigma_{\rm disc}(\sin(2\pi\beta\HS))$ (note that, for
$-1/4\beta\leq x<0$, $\sin(2\pi\beta x)$ is strictly monotone increasing).
In this way we can show that, in the present case, the assumption in Theorem \ref{thm-uwt-HS}
holds. Thus $\sin(2\pi\beta \HS)$ has  an ultra-weak time  operator  with a dense ultra-weak CCR-domain. 
In particular,  
$\sin(2\pi\beta H_{\rm hyd})$  has  an ultra-weak time  operator  with a dense ultra-weak CCR-domain. 
}\end{example}

In the same manner as above, one can find  many concrete functions $f$ such that
$f(\HS)$ has an ultra-weak time  operator  with a dense ultra-weak CCR-domain.

\subsection*{Acknowledgments}
F. H. thanks Atsushi Inoue and Konrad Schm\"udgen for helpful discussions in a workshop held in Fukuoka University at March of 2014,  and Jun Uchiyama and Erik Skibsted for sending papers \cite{uch87, fs04} to him,  respectively.   
He also thanks Aarhus University in Denmark and Rennes I University in France for kind hospitality. This work was partially done at these universities.

\end{document}